\documentclass{aa}  
\usepackage{graphicx}
\usepackage{amssymb}
\usepackage{amsmath}
\usepackage{txfonts}
\usepackage{url}
\usepackage{natbib}
\usepackage{amssymb}
\usepackage{amsmath}
\begin{document}
   \title{CARS: The CFHTLS-Archive-Research Survey \\ II. Weighing
     dark matter halos of Lyman-break galaxies at $z$=3-5\thanks{Based
       on observations obtained with MegaPrime/MegaCam, a joint
       project of CFHT and CEA/DAPNIA, at the Canada-France-Hawaii
       Telescope (CFHT) which is operated by the National Research
       Council (NRC) of Canada, the Institut National des Sciences de
       l'Univers of the Centre National de la Recherche Scientifique
       (CNRS) of France, and the University of Hawaii. This work is
       based in part on data products produced at TERAPIX and the
       Canadian Astronomy Data Centre as part of the
       Canada-France-Hawaii Telescope Legacy Survey, a collaborative
       project of NRC and CNRS. Based on zCOSMOS and VVDS observations
       carried out using the Very Large Telescope at the ESO Paranal
       Observatory under Programme IDs: LP175.A-0839 and
       070.A-9007. Based on DEEP2 observations obtained at the
       W. M. Keck Observatory. Funding for the DEEP2 survey has been
       provided by NSF grants AST95-09298, AST-0071048, AST-0071198,
       AST-0507428, and AST-0507483 as well as NASA LTSA grant
       NNG04GC89G.  }} \titlerunning{Weighing dark matter halos of
     LBGs}

   \subtitle{}

   \author{H. Hildebrandt\inst{1}
     \and J. Pielorz\inst{2}
     \and T. Erben\inst{2} 
     \and L. van Waerbeke\inst{3}
     \and P. Simon\inst{4}
     \and P. Capak\inst{5}
   }

   \offprints{H. Hildebrandt}

   \institute{Leiden Observatory, Leiden University, Niels Bohrweg 2,
     2333CA Leiden, The
     Netherlands\\ \email{hendrik@strw.leidenuniv.nl} \and
     Argelander-Institut f\"ur Astronomie, Auf dem H\"ugel 71, 53121
     Bonn, Germany\\ \and University of British Columbia, Department
     of Physics and Astronomy, 6224 Agricultural Road, Vancouver,
     B.C. V6T 1Z1, Canada\\ \and The Scottish Universities Physics
     Alliance (SUPA), Institute for Astronomy, School of Physics,
     University of Edinburgh, Royal Observatory, Blackford Hill,
     Edinburgh EH9 3HJ, UK\\ \and Spitzer Science Center, 314-6,
     California Institute of Technology, 1201 E.  California Blvd,
     Pasadena, CA, 91125
}

   \date{Received ; accepted }

  \abstract {}{We measure the clustering properties for a large
    samples of $u$- ($z\sim3$), $g$- ($z\sim4$), and $r$- ($z\sim5$)
    dropouts from the Canada-France-Hawaii Telescope Legacy Survey
    (CFHTLS) Deep fields.}{Photometric redshift distributions along
    with simulations allow us to de-project the angular correlation
    measurements and estimate physical quantities such as the
    correlation length, halo mass, galaxy bias, and halo occupation as
    a function of UV luminosity.}{For the first time we detect a
    significant one-halo term in the correlation function at $z\sim5$.
    The comoving correlation lengths and halo masses of LBGs are found
    to decrease with decreasing rest-frame UV-luminosity.  No
    significant redshift evolution is found in either quantity. The
    typical halo mass hosting an LBG is $M\ga10^{12}h^{-1}M_\odot$ and
    the halos are typically occupied by less than one galaxy.
    Clustering segregation with UV luminosity is clearly observed in
    the dropout samples, however redshift evolution cannot clearly be
    disentangled from systematic uncertainties introduced by the
    redshift distributions. We study a range of possible redshift
    distributions to illustrate the effect of this choice.
    Spectroscopy of representative subsamples is required to make
    high-accuracy absolute measurements of high-$z$ halo masses.}{}


   \maketitle
%
%________________________________________________________________

\section{Introduction}
More than a decade after its first applications the Lyman-break
technique \citep{1996ApJ...462L..17S, 2002ARA&A..40..579G} is still
the most widely used method to study galaxies in the high-redshift universe. 
Recently, our knowledge of these early cosmic epochs has
vastly increased providing an opportunity to test theories of
galaxy formation. One crucial aspect of these models is the connection
between the precisely-simulated dark matter (DM) structures and the
properties of the galaxies forming inside of them. Contemporary models
predict that galaxy properties are tightly connected to the underlying dark
matter halo hosting the galaxy, in particular the halo mass.

N-body simulations of the DM distribution and its evolution predict
the clustering properties of DM halos which are strongly correlated to
the halo mass.  Since galaxies are assumed to follow the DM, measuring
the clustering properties of galaxies with different properties,
e.g. different luminosities, allows us to link these properties to the
mass of the underlying dark matter halos.  However, galaxies are
generally a biased tracer of the DM halos, so the link is indirect.

Recently, several studies
\citep{2005ApJ...635L.117O,2006ApJ...642...63L,2007A&A...462..865H}
have detected deviations from the large scale power-law behaviour on
small scales in the correlation functions of Lyman-break galaxies
(LBGs).  These deviations are explained by the halo model \citep[see
  e.g. ][]{2000MNRAS.318..203S,2002PhR...372....1C} as galaxy pairs in
the same halo (one-halo term) allowing for a more accurate
determination of the mean halo masses and also a measure of the mean
number of galaxies hosted by a typical halo.

Here, we present a precision study of clustering properties for three
samples of LBGs at $z\sim3$, $z\sim4$, and $z\sim5$.  The sample is
coherently selected from one dataset, the CFHTLS Deep Survey, which
controls for systematic errors that affect measurements of clustering
evolution, such as the definition of selection criteria, the role of
different filter sets, masking, and the redshift-distributions.

The data used for this study, the data reduction, and the catalogue
creation are presented in Sect.~\ref{sec:data}. The selection of the
LBG samples together with their basic properties is covered in
Sect.~\ref{sec:samples}. In Sect.~\ref{sec:clustering} we describe how
we measure the correlation function of these galaxies and present the
results of model fits to the data. These results are discussed and
concluding remarks are given in Sect.~\ref{sec:conclusions}.

Throughout the paper we use AB magnitudes and we assume a $\Lambda$CDM
concordance cosmological model ($\Omega_\Lambda=0.7$,
$\Omega_\mathrm{m}=0.3$, $\sigma_8=0.9$,
$H_0=100h^{-1}\mathrm{\frac{km}{s\cdot Mpc}}$, $h=0.7$).

\section{The data}
\label{sec:data}
\subsection{Overview and reduction}
For this work we consider publicly available and \emph{Elixir}
\citep{2004PASP..116..449M} pre-processed data from the CFHTLS-Deep
($ugriz$-bands) and COSMOS survey ($u$-band).  We retrieve from the
CADC archive relevant images available at 21/07/2008.

The data were observed with MEGACAM@CFHT and reduced with the THELI
imaging reduction pipeline \citep{2005AN....326..432E}. The properties
of the reduced images are summarised in Table~\ref{tab:images} ( the
limiting magnitudes reported in the table are 1-$\sigma$ limits in a
circular aperture with a diameter of twice the seeing FWHM).  The data
reduction and catalogue creation are carried out identically to the
one of the CFHTLS-Wide survey described in
\cite{2009A&A...493.1197E}. Here, we quickly summarise the most
important reduction- and catalogue creation steps:

\begin{enumerate}
  \item Retrieval of all available pre-reduced exposures from the CADC
    archive.
  \item Quality control of all 36 chips of each retrieved exposure to
    identify bad ones (e.g. chips with a large fraction of saturated
    pixels, etc.).
  \item Identification of satellite tracks with Hough transform
    techniques \citep{2001misk.conf..595V}.
  \item Creation of weight images for each chip including bad pixels
    and the satellite track masks.
  \item Extraction of catalogues with \emph{SExtractor}
    \citep{1996A&AS..117..393B} for astrometric calibration.
  \item Astrometric calibration of the $i$-band images with the
    \emph{Astrometrix} \citep{2002ASTROMETRIX, 2003A&A...410...17M}
    software and the USNO-B1 astrometric catalogue
    \citep{2003AJ....125..984M} as a reference. After the $i$-band
    image has been stacked (step 8), the other images are calibrated
    astrometrically with a catalogue extracted from that co-added
    $i$-band image as a reference. This is done to ensure an optimal
    alignment of the images in the different bands.
  \item Relative photometric calibration with the \emph{Photometrix}
    software and absolute photometric calibration with the methods
    described in \cite{2006A&A...452.1121H}.
  \item Co-addition with the \emph{SWarp} software \citep{2003SWarp}.
  \item PSF equalisation by degrading all bands of one particular
    pointing to the seeing of the image with the worst seeing. This is
    done by convolution with appropriate Gaussian kernels.
  \item Creation of limiting magnitude maps from the noise properties
    of the co-added images.
  \item \emph{SExtractor} runs in dual-image mode with the unconvolved
    (i.e. not degraded to a worse seeing) $i$-band image as the
    detection image and the convolved images in all five bands as
    measuring images to extract a multi-colour catalogue.
  \item Automated masking of the image stacks to mask out low-S/N
    regions and regions affected by stellar diffraction spikes and
    stellar light halos as well as asteroid tracks.
\end{enumerate}

\begin{table*}
\caption{Properties of the images used in this study.}
\label{tab:images}
\centering
\begin{tabular}{ccccccrrrr}
\hline
\hline
Field & $u_{\rm lim}$ & $g_{\rm lim}$ & $r_{\rm lim}$ & $i_{\rm lim}$ & $z_{\rm lim}$ & No. of $u$-dr. & No. of $g$-dr. & No. of $r$-dr. & area\\
      & [$\rm mag_{AB}$] & [$\rm mag_{AB}$] & [$\rm mag_{AB}$] & [$\rm mag_{AB}$] & [$\rm mag_{AB}$] &  &  &  & [sq. arcmin.]\\
\hline
D1 & 29.4 & 29.8 & 29.6 & 29.5 & 28.2 & 11\,206 &    9410 &    2552 &    2989\\
D2 & 29.1 & 29.5 & 29.5 & 29.4 & 28.3 &    5652 &    9277 &    2742 &    2598\\
D3 & 29.3 & 29.8 & 29.7 & 29.5 & 28.2 & 10\,957 & 10\,118 &    2638 &    2968\\
D4 & 29.2 & 29.7 & 29.6 & 29.3 & 28.2 &    6403 &    7421 &    2550 &    2856\\
\hline
total                       & & & & & & 34\,218 & 36\,226 & 10\,482 & 11\,411\\
\end{tabular}
\end{table*}

\subsection{Photometric redshifts}
In contrast to the method applied in \cite{2009A&A...493.1197E} we modify the
estimation of photometric redshifts (photo-$z$) in the following
points:

\begin{itemize}
\item Galactic extinction is included on an object basis from the
  beginning. In \cite{2009A&A...493.1197E} we relied on the average removal of
  extinction effects by means of zeropoint (ZP) re-calibration on
  fields with spectroscopic coverage. This step decreases the scatter
  of the photo-$z$'s slightly.
\item We substitute the template set (the four empirical templates
  from \citeauthor{1980ApJS...43..393C}~\citeyear{1980ApJS...43..393C}
  plus two starburst templates from
  \citeauthor{1996ApJ...467...38K}~\citeyear{1996ApJ...467...38K}) by
  a re-calibrated version of these templates from
  \cite{2004_Capak_PhDT}. The same template set was also used by
  \cite{2007ApJS..172..117M} and is based on the technique described
  in \cite{1999ASPC..191...19B}. The new template set has an impact on
  the accuracy of the photo-$z$'s at higher redshift. The systematic
  underestimation of the redshifts, seen in \cite{2009A&A...493.1197E} on the
  CFHTLS-Wide data, vanishes.
\item We re-run the ZP re-calibration matching the colours of objects
  with spectroscopic redshifts with the colours of the best-fit
  template at that particular redshift \citep[see ][ for details of
    the procedure]{2009A&A...493.1197E}. For the D4 field, which does
  not have any spectroscopic overlap, we apply the ZP offsets found on
  the D1 field, which has the largest overlap with a spectroscopic
  catalogue and thus potentially the most accurate re-calibration.
\item The priors used in the \emph{BPZ} code
  \citep{2000ApJ...536..571B} are modified to increase the probability
  of objects being assigned a low redshift ($z \la 0.2$). In this way
  we are also able to remove the systematic overestimation of
  redshifts for $z\la0.2$ as reported in \cite{2009A&A...493.1197E}.
\end{itemize}

By these measures a tilt in the photo-$z$ vs. spec-$z$ comparison, as
reported in \cite{2009A&A...493.1197E} can be avoided.

\begin{figure}
\centering
\resizebox{\hsize}{!}{\includegraphics{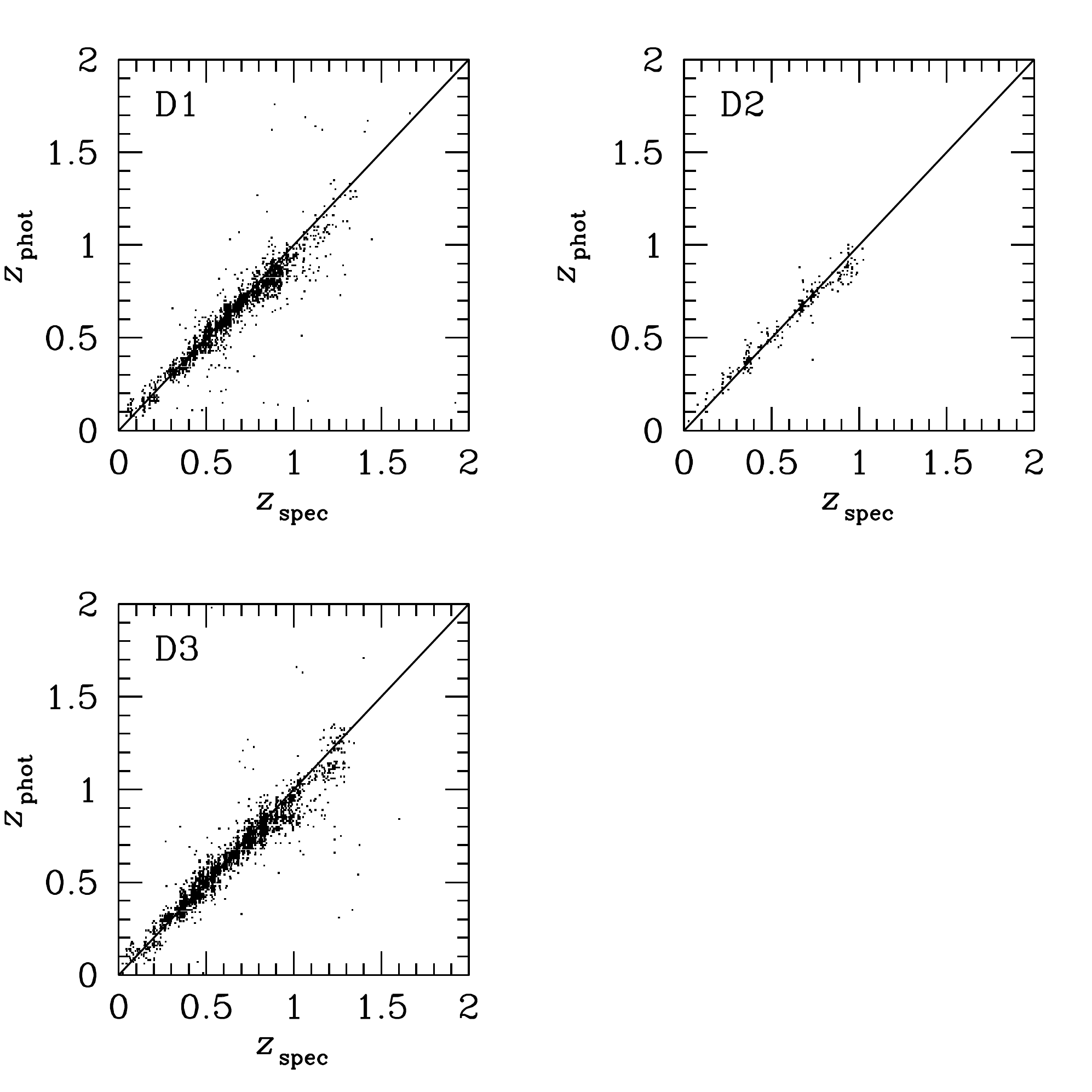}}
\caption{Comparison of the \emph{BPZ} photo-$z$'s to spectroscopic
  redshifts from the VVDS, zCOSMOS, and DEEP2 (2291, 232, and 2705
  objects with good spectroscopic flags and $\mathrm{ODDS}>0.9$ in the
  D1, D2, and D3 fields, respectively).}
\label{fig:zz}
\end{figure}

With thousands of public spectroscopic redshifts from the VIMOS VLT
Deep Survey (VVDS) on the D1 field \citep{2005A&A...439..845L}, the
zCOSMOS survey on the D2 field \citep{2007ApJS..172...70L}, and the
DEEP2 survey on the D3 field \citep{2007ApJ...660L...1D}, we test the
performance of our photo-$z$'s. For a safe sample
\citep[$\mathrm{ODDS}>0.9$; see also ][]{2000ApJ...536..571B,
  2008A&A...480..703H, 2009A&A...493.1197E} in the magnitude range
$17<i<24$ we find a scatter of $\sigma=0.033$ of the quantity $\Delta
z=\left(z_{\rm phot}-z_{\rm spec}\right)/\left(1+z_{\rm spec}\right)$
after rejecting 1.6\% of outliers (objects with $\Delta z > 0.15$) and
no significant bias. Figure~\ref{fig:zz} shows a comparison of the
spectroscopic redshifts with our photo-$z$'s on the three fields. Note
that these accurate low-redshift photo-$z$'s for relatively bright
objects do not imply directly a similar photo-$z$ performance at
higher redshifts and fainter magnitudes. This results should be
regarded more as a quality check for the data reduction and the
multi-colour photometry.

Furthermore, we also carried out a cross-correlation analysis of
galaxies in photo-$z$ bins. The results can be found in the
appendix~\ref{app:cross}.

\section{The LBG samples}
\label{sec:samples}
\subsection{Simulations}
We simulate a colour catalogue of galaxies based on templates from the
library of \cite{1993ApJ...405..538B} - in the same way as presented
in \cite{2007A&A...462..865H} for the ESO Deep Public Survey
(DPS):

\begin{itemize}
  \item We take the observed $i$-band number counts as a reference for
    creating the simulated colour catalogue.
  \item $i$-band magnitude-dependent and spectral-type-dependent
    redshift distributions are extracted from the \emph{BPZ}
    code. These are based on catalogues extracted from the Hubble Deep
    Field North \citep{1999ApJ...513...34F} and the Canada-France
    Redshift Survey \citep{1995ApJ...455..108L}.
  \item A huge multi-colour catalogue is simulated with the
    \emph{Hyperz} \citep{2000A&A...363..476B} photo-$z$ code evenly
    distributed in $i$-band magnitude, redshift, and spectral-type
    \citep[eight different star-formation-histories applied to the
      library of ][]{1993ApJ...405..538B}.
  \item From this huge homogeneous catalogue we randomly pick objects
    for a given magnitude, type, and redshift bin. The numbers in a
    particular bin are scaled by the $i$-band number counts and the
    magnitude- and type-dependent redshift distributions mentioned
    above.
\end{itemize}

The colour catalogue simulated in this way shows very similar
colour-colour diagrams as the observed catalogue \citep[see also
][]{2007A&A...462..865H} and is supposed to represent a realistic mix
of galaxy types and low- and high-redshift objects.

In order to study the effect of the choice of the template set we
re-run the simulations with the templates from
\cite{2006ApJ...652...85M} \citep[see also ][]{2005MNRAS.362..799M} as
a basis.\footnote{ These templates can be retrieved from
  \url{http://www.icg.port.ac.uk/~maraston/hyperz-templates/}} In
general, the results are very similar to the ones with the template
set from \cite{1993ApJ...405..538B}. The scientific results in this
study are hardly affected by this choice. Nevertheless, we show
results for both template sets in the remainder of the paper.

Furthermore, we simulate the colours of stars in our fields from the
TRILEGAL galactic model \citep{2005A&A...436..895G} taking into
account the galactic coordinates of our survey fields and the depths
of the images \citep[see also ][]{2007A&A...462..865H}.

\subsection{Selection}
\begin{figure*}
\centering
\includegraphics[width=5.5cm]{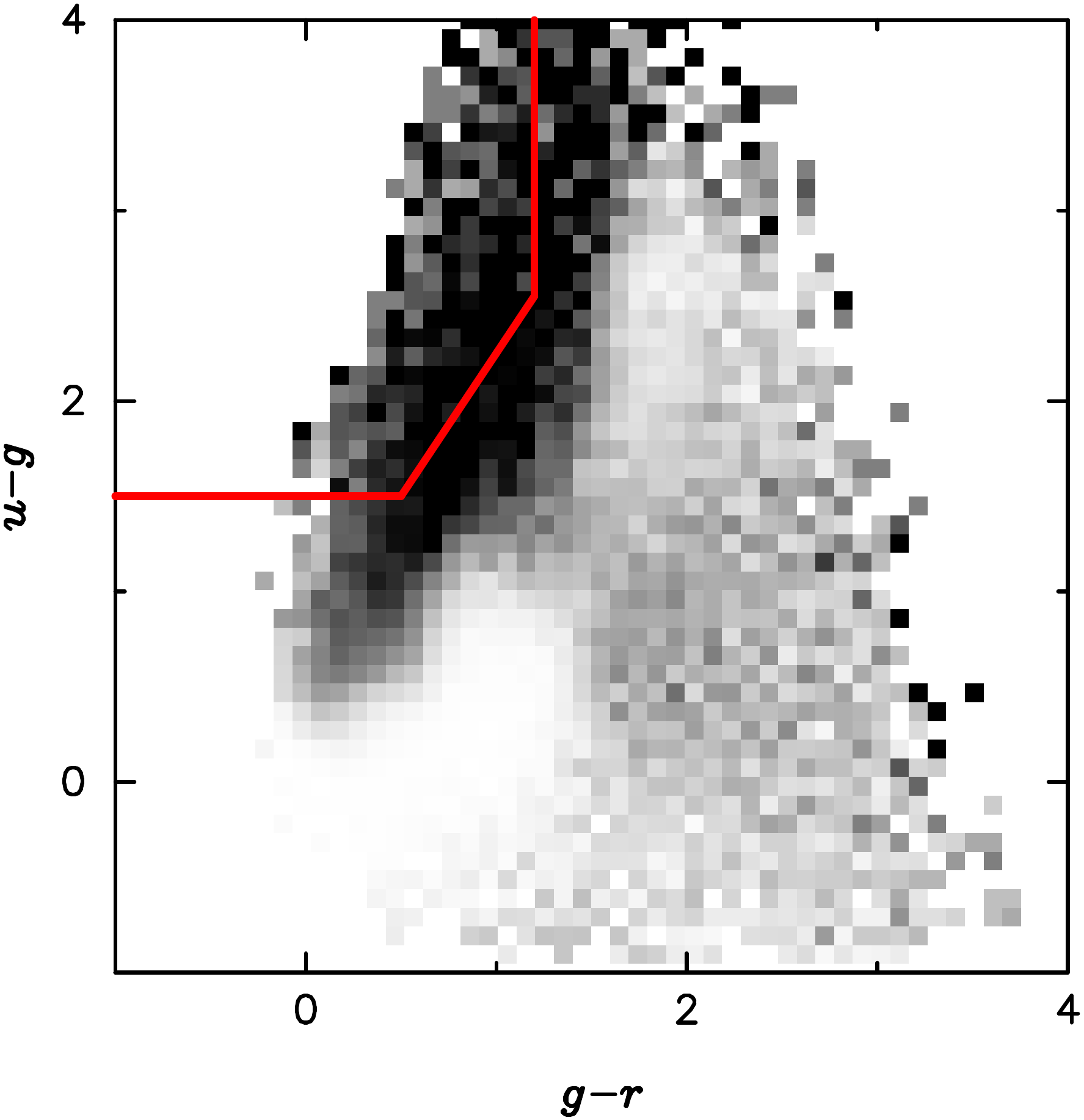}
\includegraphics[width=5.5cm]{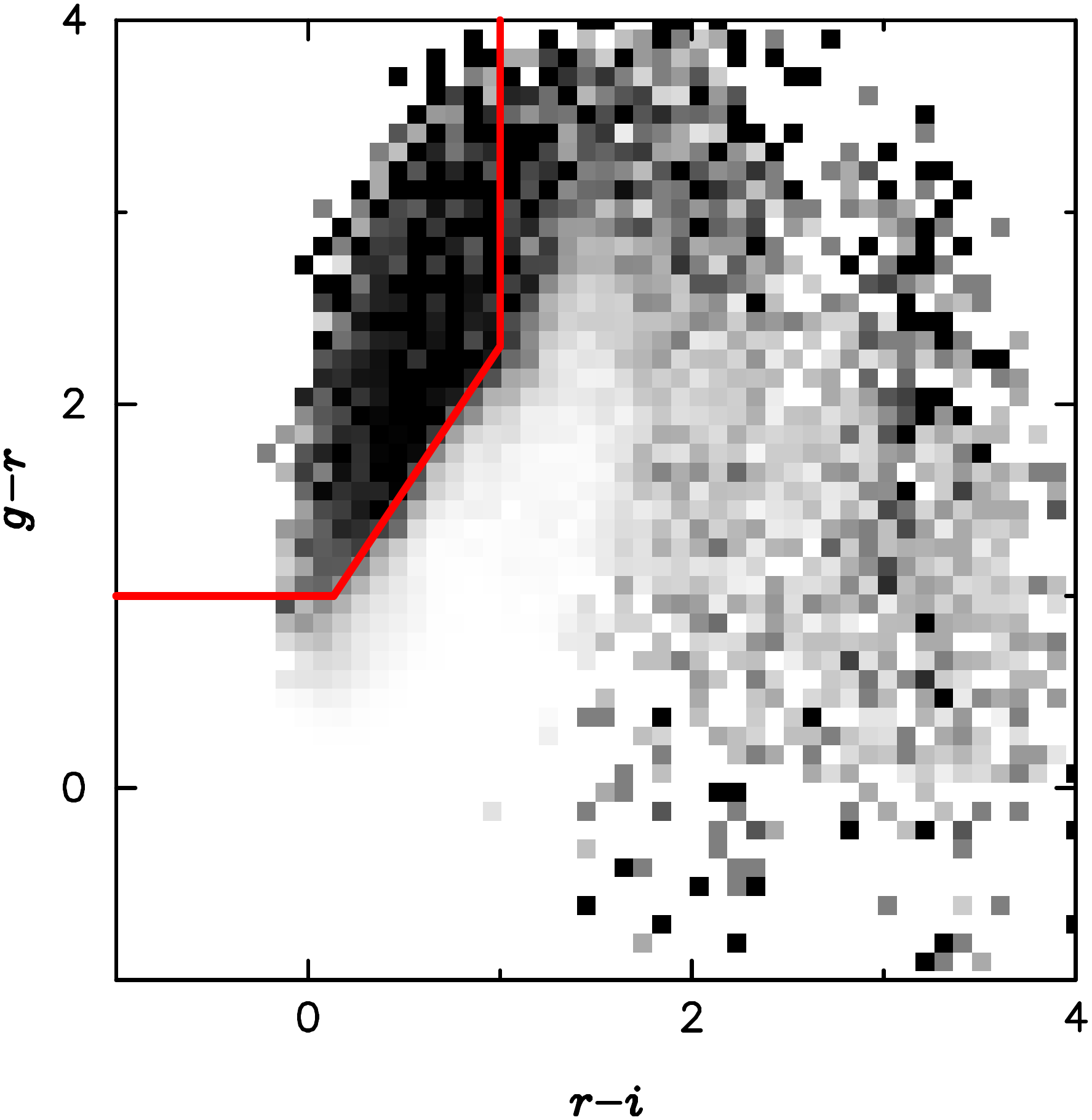}
\includegraphics[width=5.5cm]{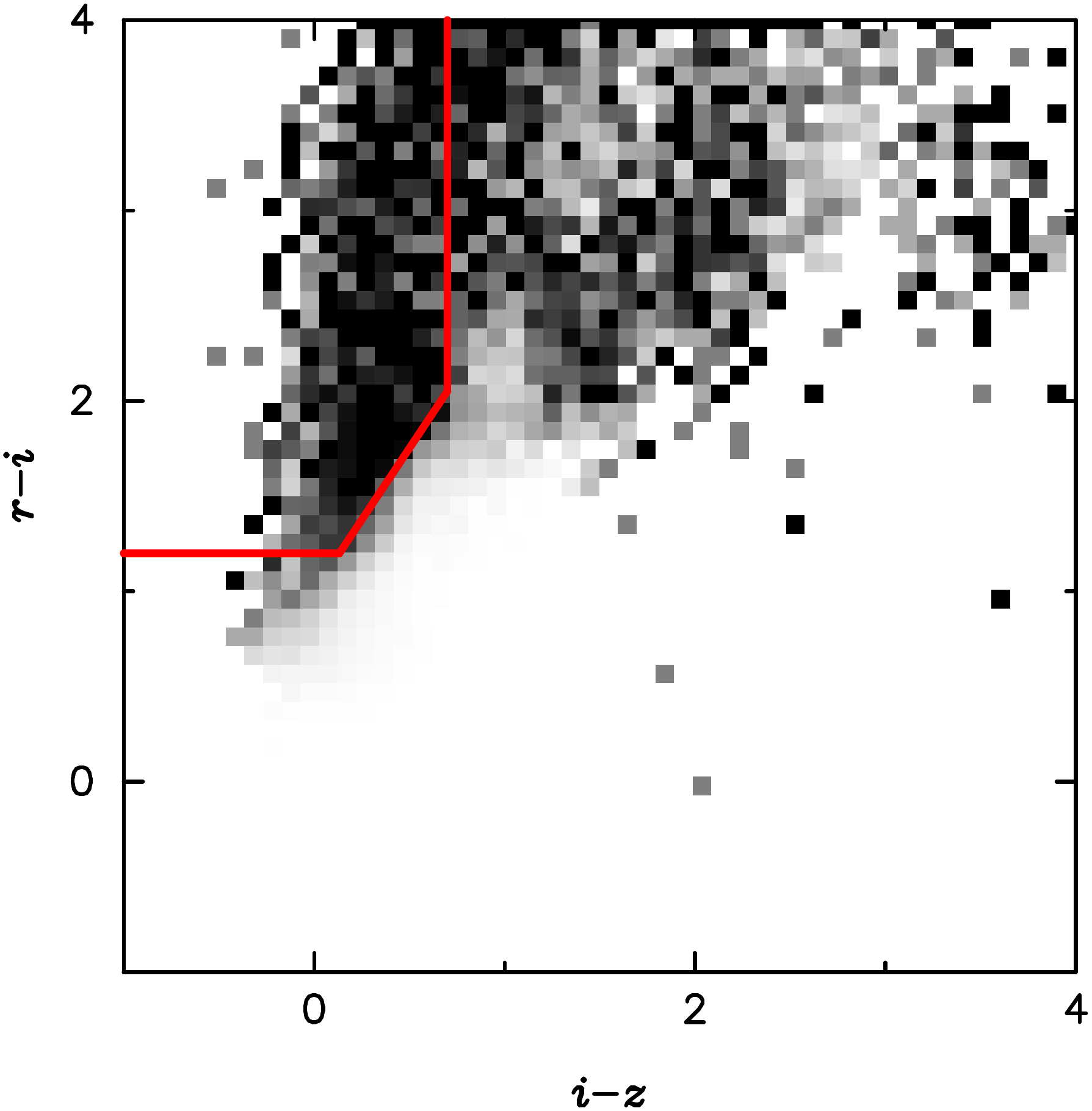}
\caption{ Colour-colour diagrams showing the fraction of objects at
  the target redshifts (\emph{left}: $2<z<4$; \emph{middle}: $3<z<5$;
  \emph{right}: $4<z<6$) as a function of colour in our simulations
  based on the templates from \cite{1993ApJ...405..538B}, with black
  areas corresponding to 100\% and white to 0\% selection
  efficiency. The solid lines represent the selection criteria adopted
  in this study.}
\label{fig:sim_col}
\end{figure*}

From these simulations we identify regions in colour-space where the
efficiency of finding high-redshift star-forming galaxies is high due
to their distinctive colours produced by the Lyman-break and where the
contamination from low-redshift interlopers and stars is low. We do
this in the $ugr$, $gri$, and $riz$ colour-spaces to select
$u$-dropouts, $g$-dropouts, and $r$-dropouts, respectively. In
Fig.~\ref{fig:sim_col} the fraction of objects in the desired redshift
range ($u$-dropouts: $2<z<4$; $g$-dropouts: $3<z<5$; $r$-dropouts:
$4<z<6$) is plotted as a function of colour. In order to steer away
from the stellar locus, not the whole high-efficiency area in the
$u-g$ vs. $g-r$ colour-space can be used.

\begin{figure*}
\centering
\includegraphics[width=5.5cm]{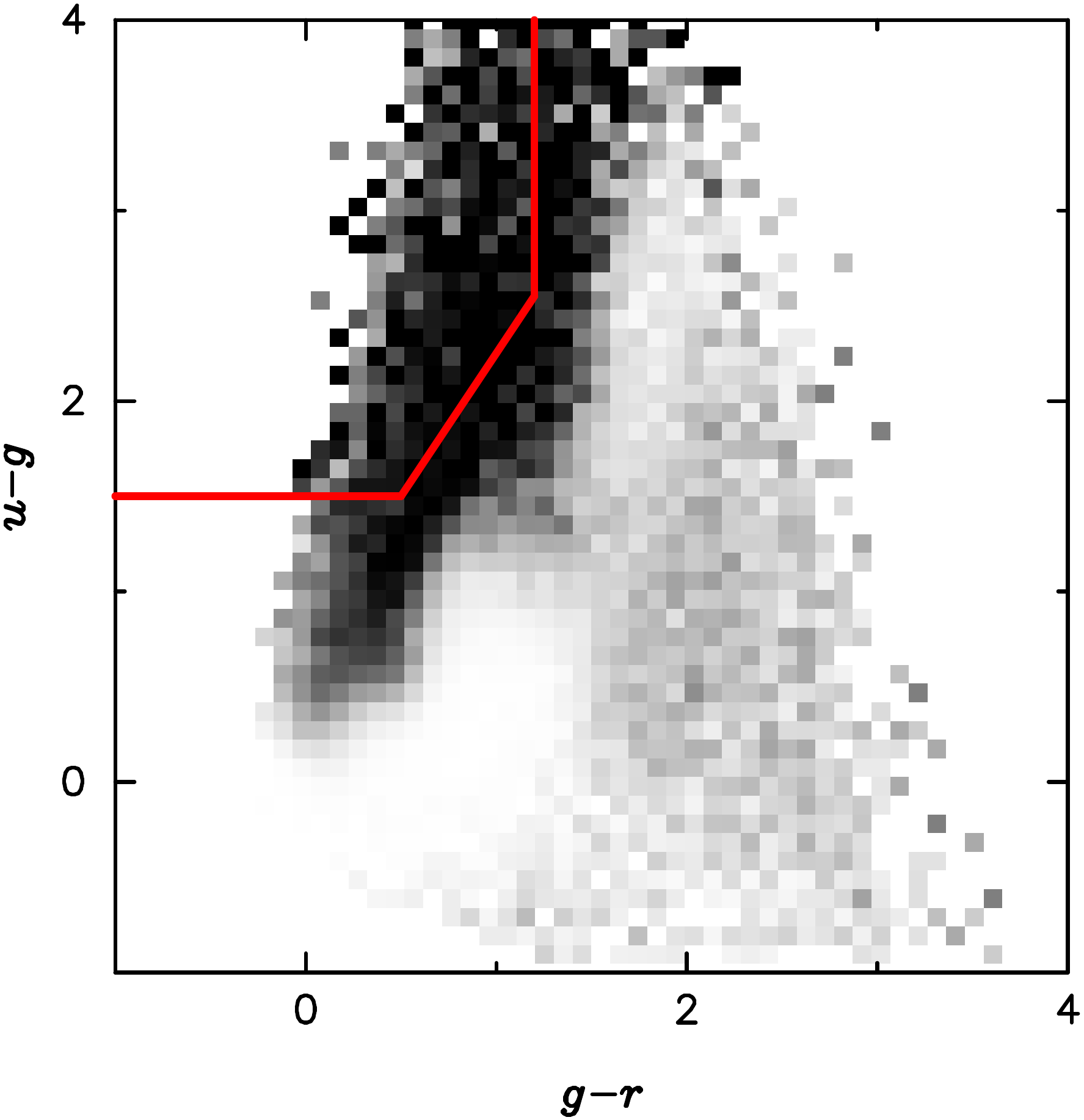}
\includegraphics[width=5.5cm]{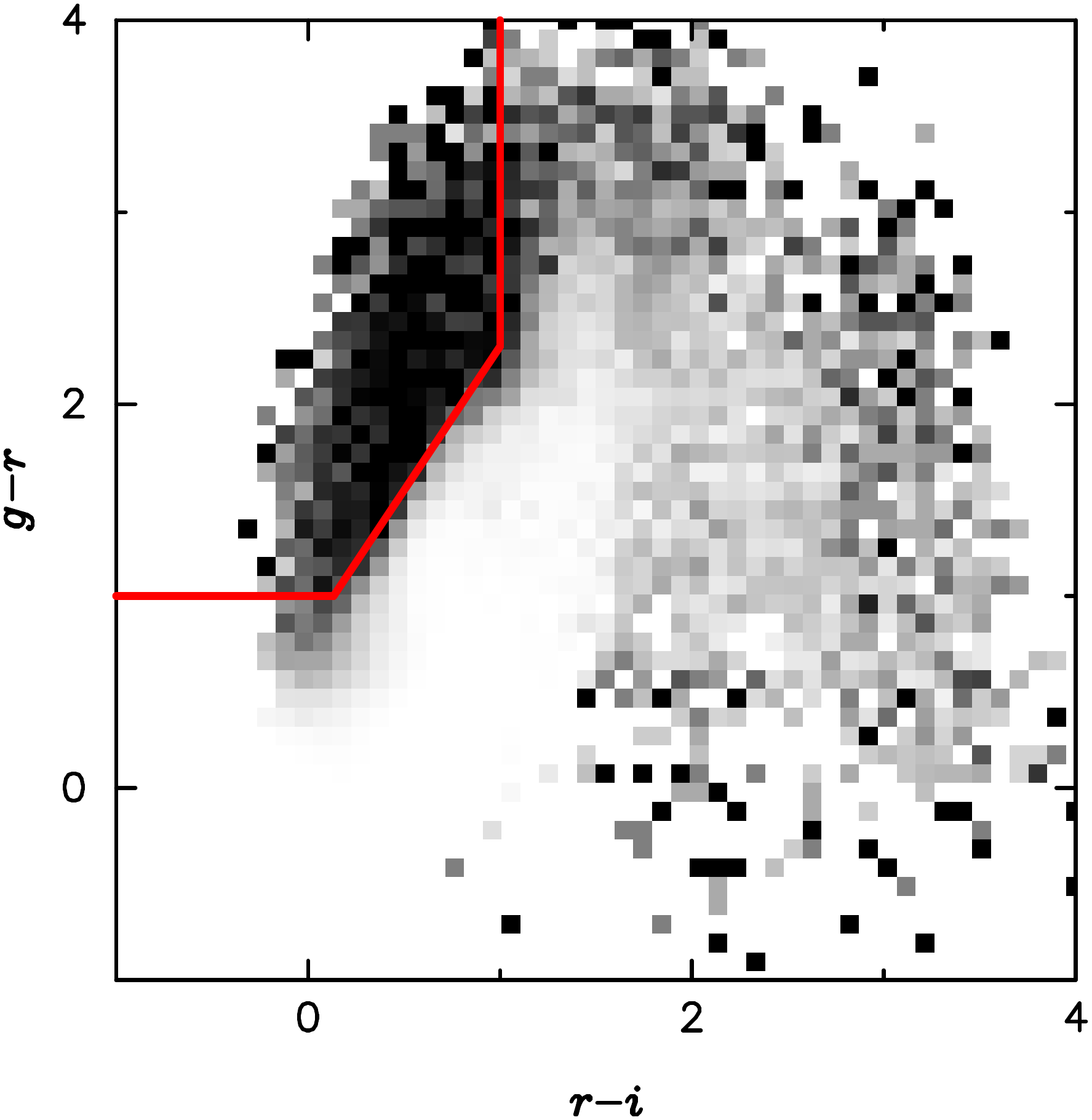}
\includegraphics[width=5.5cm]{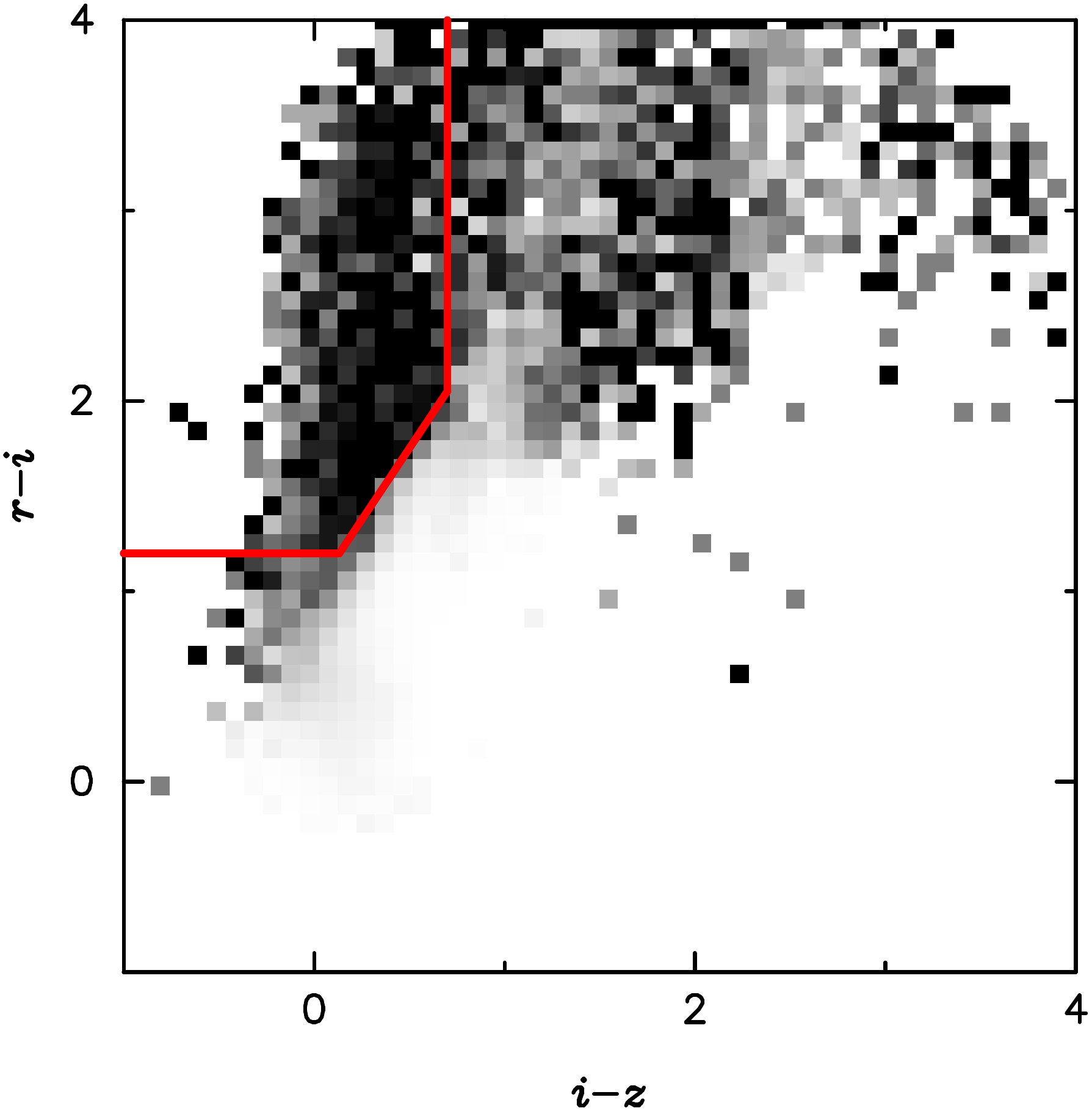}
\caption{Same as Fig.~\ref{fig:sim_col} but based on the templates
  from \cite{2006ApJ...652...85M} showing that the choice of the
  template set does not make any difference here.}
\label{fig:sim_col_Ma}
\end{figure*}

Our colour selection criteria are:
\begin{itemize}
\item for the $u$-dropouts: $1.5 <(u-g) \: \wedge \: -1.0 < (g-r)<1.2 \: \wedge \: 1.5\cdot(g-r) < (u-g)-0.75$\,,
\item for the $g$-dropouts: $1.0 <(g-r) \: \wedge \: -1.0 < (r-i)<1.0 \: \wedge \: 1.5\cdot(r-i) < (g-r)-0.8$\,,
\item for the $r$-dropouts: $1.2 <(r-i) \: \wedge \: -1.0 < (i-z)<0.7 \: \wedge \: 1.5\cdot(i-z) < (r-i)-1.0$\,.
\end{itemize}

Furthermore, we require all LBGs to have a \emph{SExtractor}
CLASS\_STAR parameter of $\mathrm{ CLASS\_STAR}<0.9$,\footnote{This
  cut on the compactness of the objects assures a rejection of most
  stars. The image quality of our $i$-band images is very high with a
  seeing FWHM$\la0\farcs7$. Thus we can still separate reliably most
  high-$z$ galaxies from stars.} that $g$-dropouts are not detected in
$u$, and that $r$-dropouts are neither detected in $u$ nor in $g$. In
this way we select 34\,218 $u$-dropouts, 36\,226 $g$-dropouts, and
10\,482 $r$-dropouts from the effective area of 3.17 sq. deg.,
i.e. the area after masking \citep[see ][ for a detailed description
  of the masking algorithms]{2009A&A...493.1197E}.

\subsection{Properties}
\label{sec:properties}
\begin{figure*}
\centering
\includegraphics[width=17cm]{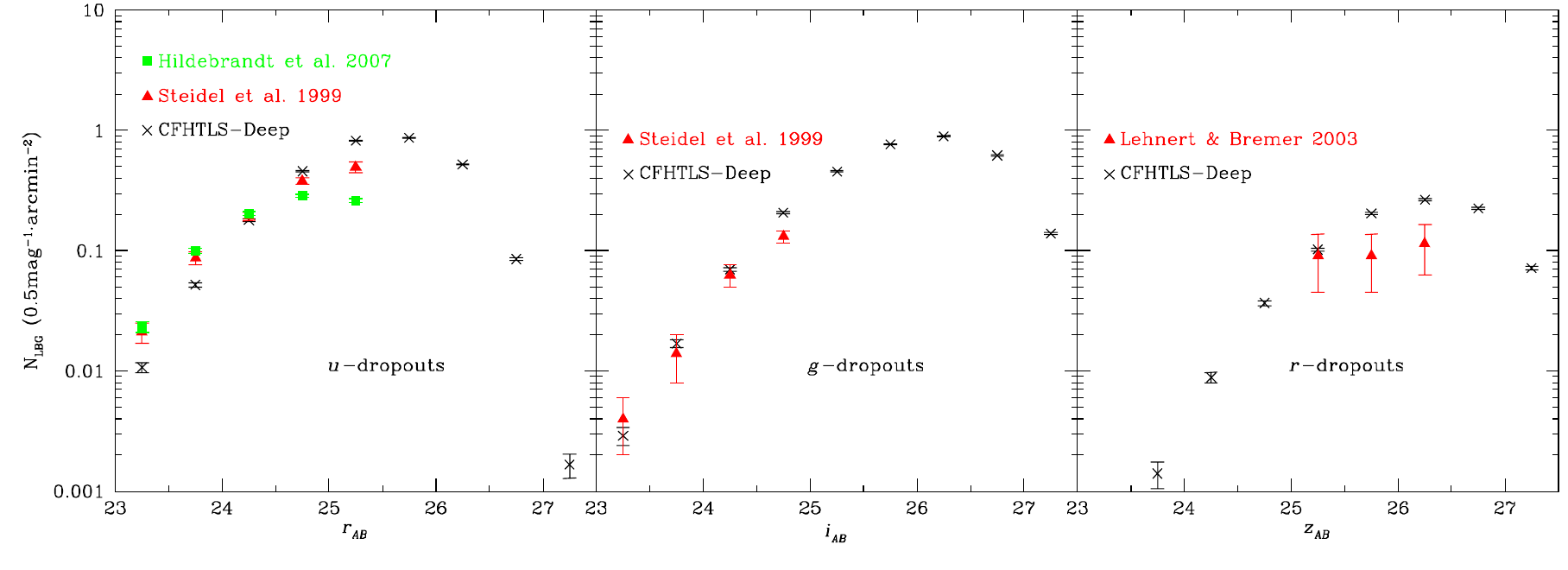}
\caption{Number counts for the three dropout samples ($u$-dropouts
  \emph{left}, $g$-dropouts \emph{middle}, and $r$-dropouts
  \emph{right}) in the bands corresponding to approximately the same
  rest-frame UV-wavelength. We compare our $u$-dropout number counts to
  the ones from \cite{1999ApJ...519....1S} and
  \cite{2007A&A...462..865H} clearly showing the greater depth of the
  CFHTLS-Deep data. The $g$-dropout number counts are compared to
  \cite{1999ApJ...519....1S} being consistent for $i<24.5$ and
  suggesting some incompleteness of the older survey for fainter
  magnitudes. The 13 $R$-dropouts reported in
  \cite{2003ApJ...593..630L} show a similar number density as the
  $r$-dropouts in the current study.}
\label{fig:numbercounts}
\end{figure*}

The number counts of the three samples are plotted in
Fig.~\ref{fig:numbercounts}. They are in good agreement with previous
studies, but the superior quality of the CFHTLS-Deep data for this
kind of analysis is clearly visible, most dramatically for the
$r$-dropout sample.

The number counts suggest approximate completeness limits of $r=25.5$
for the $u$-, $i=26.0$ for the $g$-, and $z=26.5$ for the
$r$-dropouts. These completeness limits should not be regarded as
robust. We only choose them by eye. Extensive simulations are prepared
to quantify reliably the completeness as a function of
magnitude. These are important for estimates of luminosity functions
but are beyond the scope of this study, especially since the
clustering measurements are not affected strongly by some low amounts
of incompleteness.

From the simulated catalogues described above we find that the
contamination from stars and low-$z$ interlopers can be kept below the
10\% level in each magnitude bin (bin-size $\Delta\mathrm{mag}=0.5$)
by applying a bright cutoff of $r>23$ for the $u$-, $i>23.5$ for the
$g$-, and $z>24$ for the $r$-dropouts \citep[for a discussion on how
  to avoid contaminants in an $R$-dropout survey by a bright cutoff
  see also ][]{2003ApJ...593..630L}. These magnitude cuts define our
high-quality samples which consist of 17\,338 $u$-, 17\,281 $g$-, and
7038 $r$-dropouts. In the following, only LBGs in these magnitude
ranges are considered.

\subsection{Redshift distributions}

The redshift distributions of the LBG samples are of crucial
importance in the physical interpretation of the angular clustering
results presented below. There is virtually no overlap between the
secure spectroscopic samples of the VVDS, zCOSMOS, and DEEP2 and our
high-quality dropout samples. Nearly all LBG candidates are too faint
and the brighter ones do not have good spectroscopic measurements as
indicated by their spectroscopic flags.  Here, we present four
different ways to estimate the redshift distribution by other
means. All four sets of redshifts distributions, which are displayed
in Fig.~\ref{fig:z_dist}, will be used in
Sect.~\ref{sec:clustering_results} to interpret the observed
clustering.

\subsubsection{Photo-$z$ distributions from \emph{BPZ}} The
photometric redshift distributions of the selected objects, estimated
with \emph{BPZ} and filtered for a \emph{BPZ} ODDS parameter of ${\rm
  ODDS}>0.9$, show clear peaks at the targeted redshifts. The means
and RMS-widths of these peaks are $z=3.28\pm0.15$, $z=3.87\pm0.32$,
and $z=4.74\pm0.14$. While the mean redshifts agree well with
expectations, the widths of the distributions for the $u$- and the
$r$- dropouts seem too small. This may well be an effect of the
aggressive prior used by \emph{BPZ} or the empirical template set,
which is based on low-$z$ observations.

If we run \emph{BPZ} without the prior, objects with a low ODDS
parameter that were assigned a low redshift before are assigned a high
redshift. This is exactly the behaviour one expects from double peaked
redshift probability functions. For fixed ODDS the width of the
$z$-distributions does not change from the prior- to the
non-prior-setup. If the prior is switched off and additionally the cut
on ${\rm ODDS}>0.9$ is dropped the $z$-distributions change only
slightly.

It is the width of the redshift distribution rather than the mean
redshift that is important for the interpretation of the clustering
measurements (correlation lengths, galaxy bias, halo masses, and halo
occupation numbers) described below.

\subsubsection{Photo-$z$ distributions from
    \emph{Hyperz}} \label{sec:hyperz}An independent check of the
photo-$z$'s with \emph{Hyperz} and the template set based on the
library of \cite{1993ApJ...405..538B} yields the following means and
RMS-widths: $z=2.93\pm0.24$ for the $u$-, $z=3.67\pm0.31$ for the
$g$-, and $z=4.47\pm0.22$ for the $r$-dropouts. The distributions are
shown in Fig.~\ref{fig:z_dist} as well. While the \emph{Hyperz} RMS
widths seem more realistic, the mean redshift values are slightly
lower than the \emph{BPZ} ones above and also lower than the
theoretically expected ones.  In \cite{2008A&A...480..703H} we found
that for a magnitude limited galaxy sample the photo-$z$'s for this
particular combination of code and template set and estimated from a
similar filter set ($UBVRI$) are biased low at the $\sim5\%$ level
when compared to spectroscopic redshifts of the VVDS. This is most
probably due to a slight mismatch of the absolute spectrophotometric
calibration of the templates and the absolute photometric calibration
of the data. Correcting for this bias would make the mean redshifts of
the $g$- and $r$-dropouts very compatible with \emph{BPZ} leaving only
the $u$-dropouts slightly low, although it is questionable whether
this bias will remain constant out to much higher redshifts than
probed by the VVDS.

The multi-modality of the photometric redshift distributions hints at
problems in the photo-$z$ estimation. No such multi-modalities are
observed in spectroscopic surveys of LBGs \citep[see
  e.g. ][]{1999ApJ...519....1S,2003ApJ...592..728S} nor are they
expected theoretically. Thus, we regard the \emph{Hyperz} redshift
distributions as the least reliable ones.

\subsubsection{Redshift distributions from simulations}
\begin{figure*}
\centering 
\includegraphics[width=4.5cm]{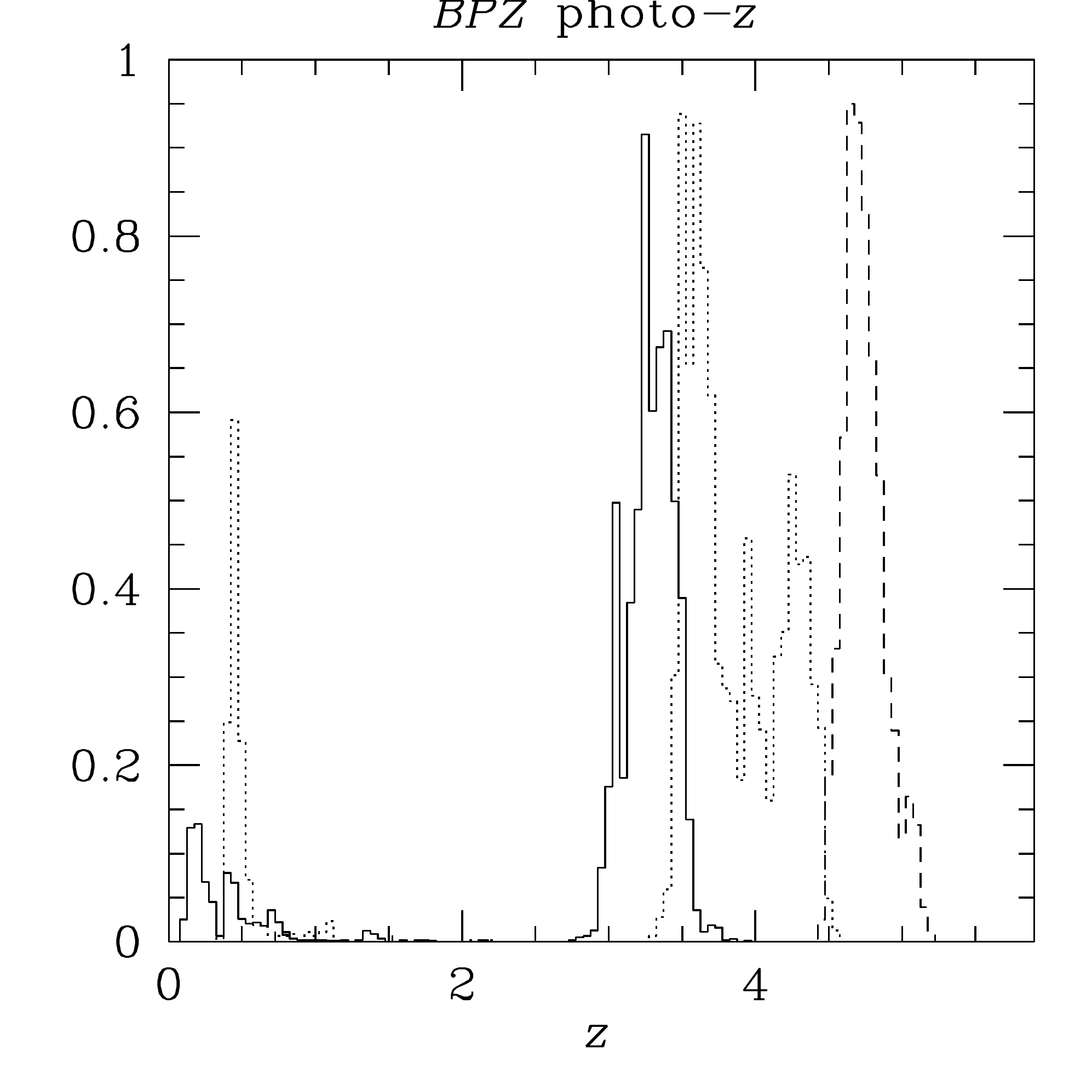}
\includegraphics[width=4.5cm]{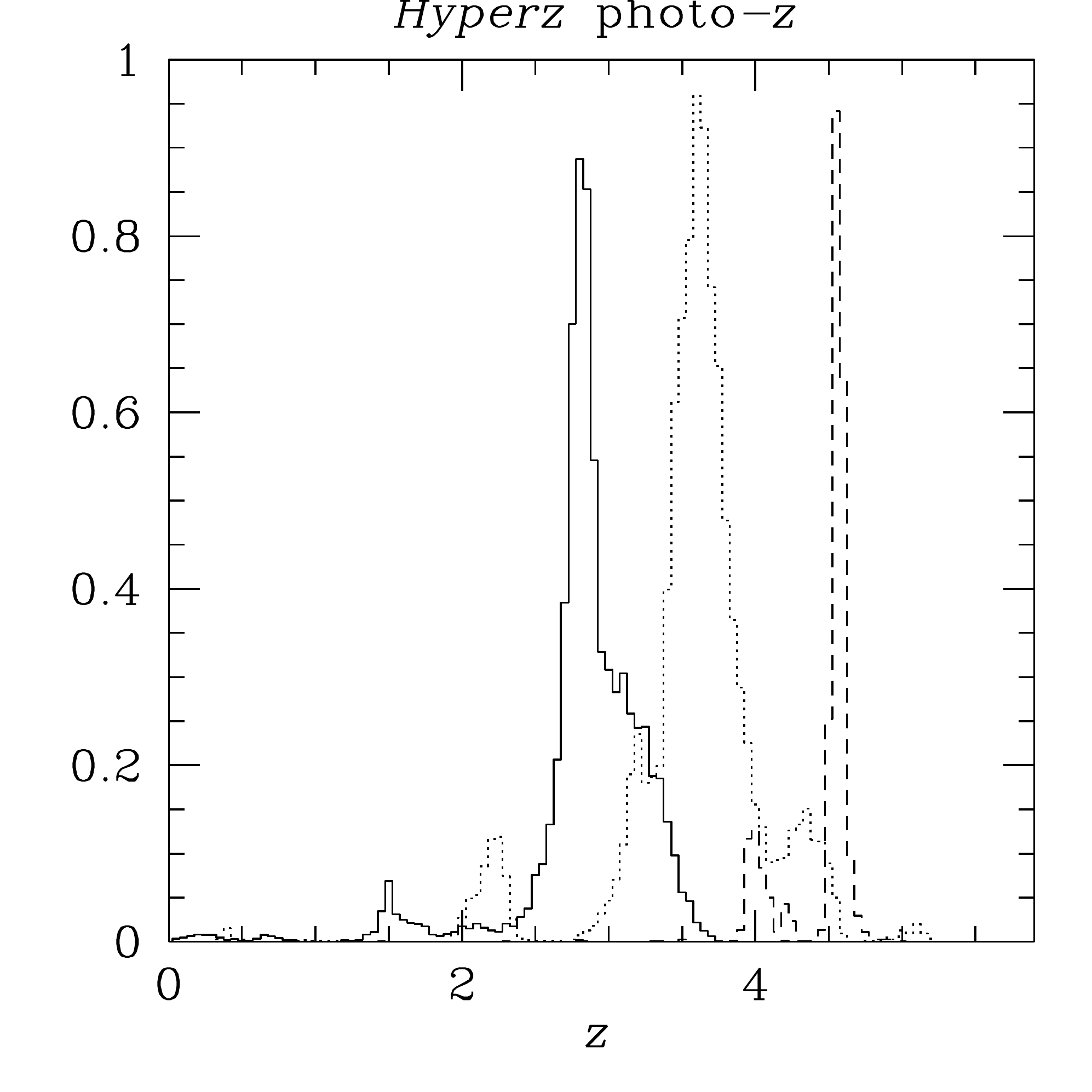}
\includegraphics[width=4.5cm]{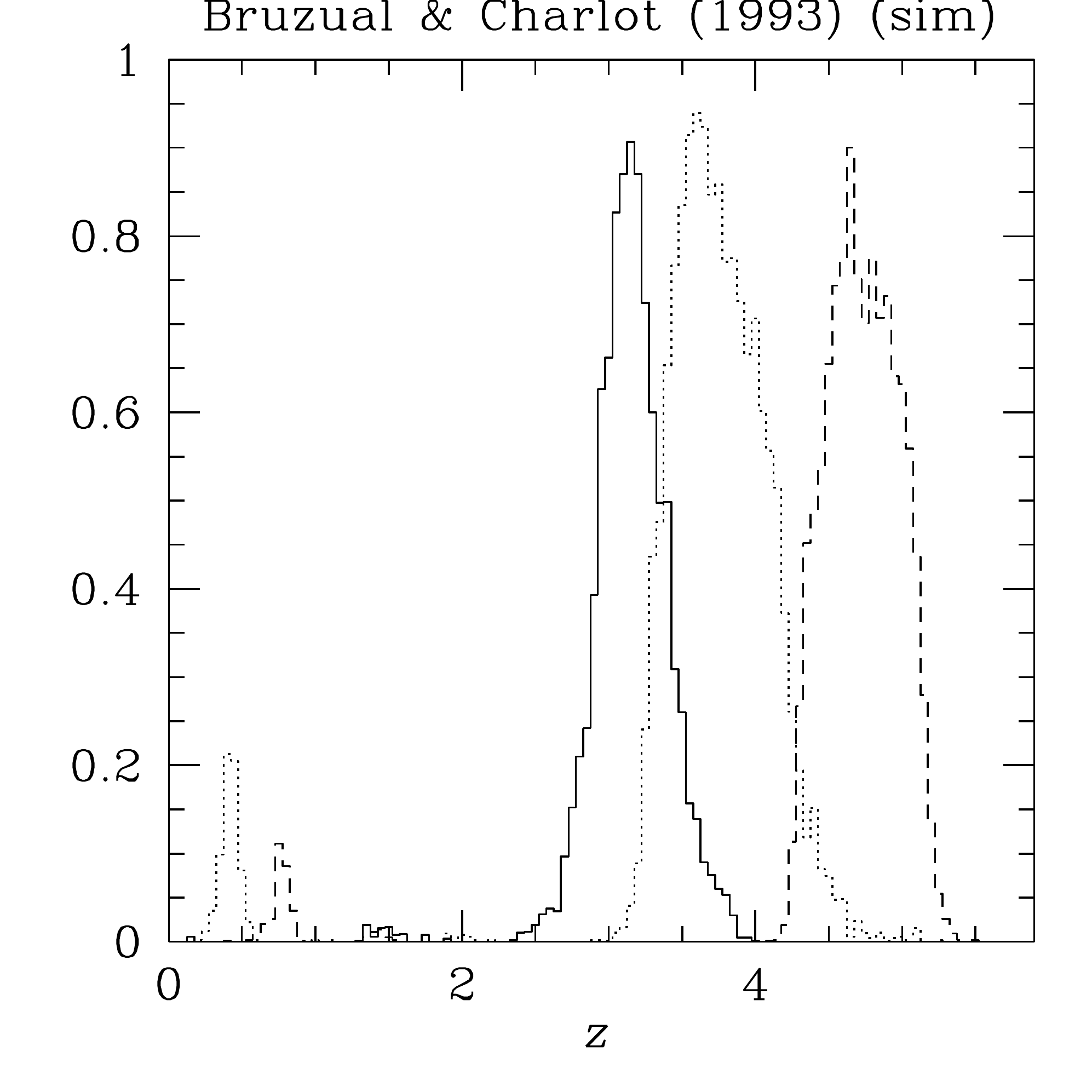}
\includegraphics[width=4.5cm]{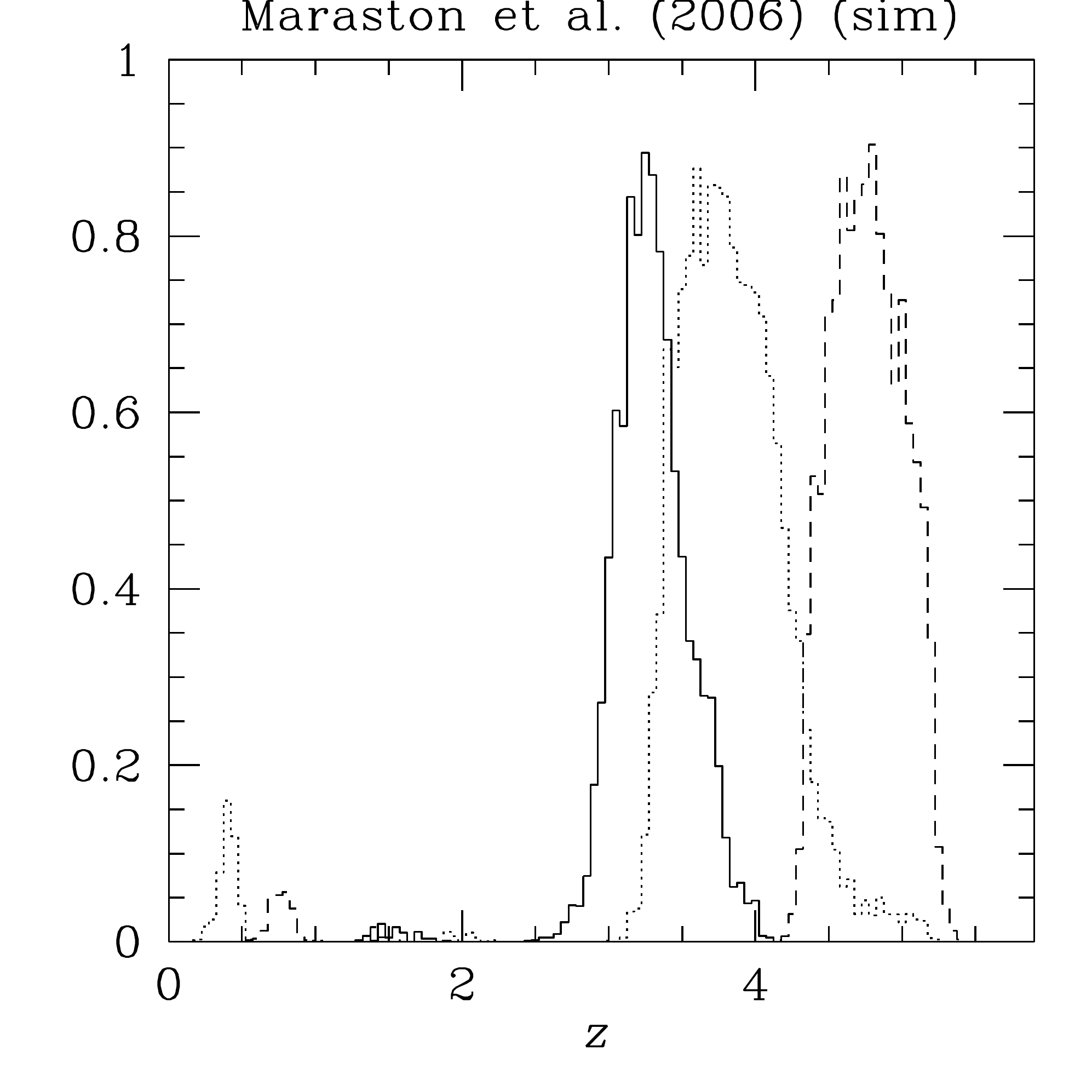}
\caption{Redshift distributions (arbitrarily normalised) of the three
  dropout samples ($u$-dropouts \emph{solid}, $g$-dropouts
  \emph{dotted}, and $r$-dropouts \emph{dashed}). The first panel
  shows the \emph{BPZ} photo-$z$ distributions, the second panel shows
  the \emph{Hyperz} photo-$z$ distributions, the third panel shows the
  distributions derived from the simulated colour catalogue based on
  \cite{1993ApJ...405..538B} templates, and the last panel shows the
  distributions derived from the simulated colour catalogue based on
  \cite{2006ApJ...652...85M} templates.}
\label{fig:z_dist}
\end{figure*}

The redshift distributions which are derived from the simulated colour
catalogues are shown in Fig.~\ref{fig:z_dist} as well. The ones based
on the templates from \cite{1993ApJ...405..538B} suggest mean
redshifts and RMS-widths of $z=3.16\pm0.22$ for the $u$-,
$z=3.76\pm0.30$ for the $g$-, and $z=4.73\pm0.24$ for the
$r$-dropouts. The simulations based on \cite{2006ApJ...652...85M}
templates yield means and widths of $z=3.30\pm0.24$, $z=3.83\pm0.32$,
and $z=4.77\pm0.24$ for the $u$-, $g$-, and $r$-dropouts,
respectively.

In the following we will refer to the four sets of redshift
distributions as the \emph{BPZ}, the \emph{Hyperz}, the
\emph{BC\_sim}, and the \emph{Maraston\_sim} distributions.

By choosing the $r$-band as the reference magnitude for the
$u$-dropouts, the $i$-band for the $g$-dropouts, and the $z$-band for
the $r$-dropouts we select galaxies at similar UV rest-frame
wavelengths (mean rest-frame wavelength of the filters is
$\sim1560\AA$ for \emph{BC\_sim}) so that the $k$-correction between
the different samples is negligible. The distance modulus between the
mean redshifts of the $u$- and the $g$-dropouts and between the mean
redshifts of the $g$- and the $r$-dropouts is $\sim0.39$mag and
$\sim0.65$mag, respectively (again for \emph{BC\_sim}).

\section{Clustering analysis}
\label{sec:clustering}
\subsection{Technique}
\begin{figure*}
%\sidecaption
%\includegraphics[width=12cm]{1042fig6.pdf}
\resizebox{\hsize}{!}{\includegraphics{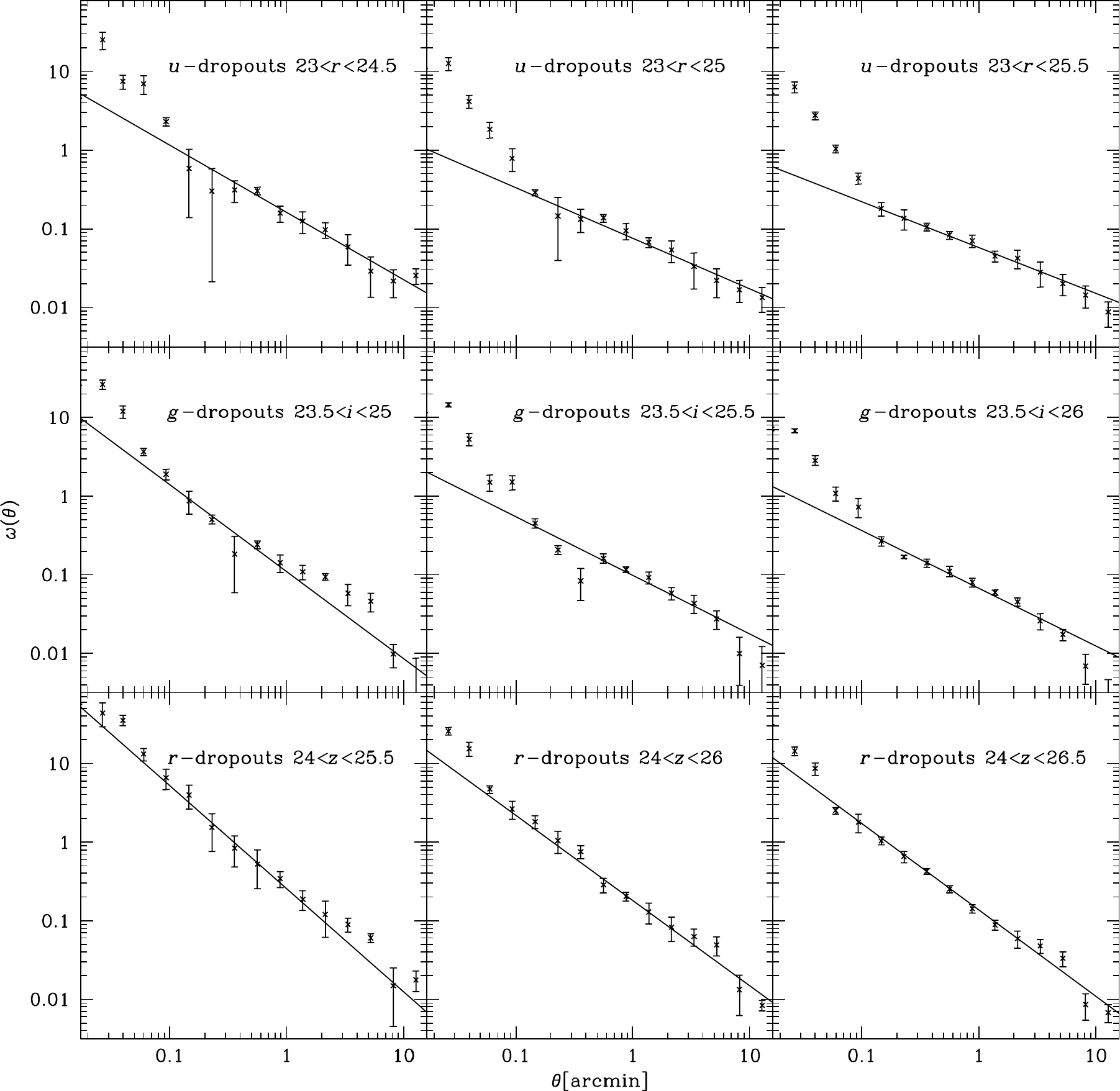}}
\caption{Angular correlation functions for the LBG samples of the
  three dropout populations ($u$-dropouts \emph{top}, $g$-dropouts
  \emph{middle}, and $r$-dropouts \emph{bottom}) for different
  limiting magnitudes (corrected for the integral constraint bias with
  the \emph{BC\_sim} redshift distributions). The errors are estimated
  from field-to-field variance over the four CFHTLS-Deep fields. The
  \emph{solid} lines represent power-law fits to the data in the
  angular region $0\farcm1<\theta<10\arcmin$. For all samples we find
  a significant deviation on small scales from this power-law
  behaviour on larger scales.}
\label{fig:omega}
\end{figure*}

We estimate the angular correlation function $\omega(\theta)$ of the
different flux-limited subsamples by applying the estimator of
\cite{1993ApJ...412...64L}. Errors are estimated from the
field-to-field variance of the four fields. We choose a common angular
binning for all samples but we check the influence of this choice (see
below). The same masks as for the object detection are used for
creating the random catalogue, which is necessary to estimate the
correlation function. See Fig.~\ref{fig:omega} for the correlation
functions of the different magnitude-limited LBG samples.

We fit a power-law, $\omega(\theta)=A\,\theta^{-\delta}$, to the
angular correlation functions in the angular region
$0\farcm1<\theta<10'$. A de-projection with Limber's
equation\footnote{See \citet{2007A&A...473..711S} for an extensive
  discussion of the accuracy of Limber's equation in this kind of
  applications. The redshift distributions of the LBGs are wide enough
  to not bias our clustering measurements significantly when applying
  Limber's equation.} \citep{1953ApJ...117..134L} involving the
redshift distributions shown above (see of Fig.~\ref{fig:z_dist})
yields the real-space correlation functions approximated by a
power-law, $\xi(r)=\left(\frac{r}{r_0}\right)^{-\gamma}$, with $r_0$
being the comoving correlation length and $\gamma=\delta+1$ being the
slope of the correlation function.  We correct iteratively for the
integral constraint bias with the method explained in
\cite{2005ApJ...619..697A}. In this way we also estimate values for
the large-scale\footnote{The fitting ranges correspond to comoving
  scales of $> 130h^{-1}{\rm kpc}$.} galaxy bias factor, $b_{\rm
  gal}$:
\begin{equation}
b_{\rm gal}=\frac{\sigma_{8,g}}{\sigma_8(z)}\,.
\end{equation}
The galaxy fluctuations in spheres of radius $8h^{-1}{\rm Mpc}$ can be
estimated from the power law fit to the correlation function in the
following way \citep{1980lssu.book.....P}:
\begin{equation}
\sigma_{8,g}=\frac{72\left(r_0/8h^{-1}{\rm
    Mpc}\right)^\gamma}{(3-\gamma)(4-\gamma)(6-\gamma)2^\gamma}\,,
\end{equation}
whereas the corresponding DM fluctuations at redshift $z$,
$\sigma_8(z)$, are calculated from theory.

\subsection{Results}
\label{sec:clustering_results}
\begin{figure*}
\centering
\resizebox{\hsize}{!}{\includegraphics{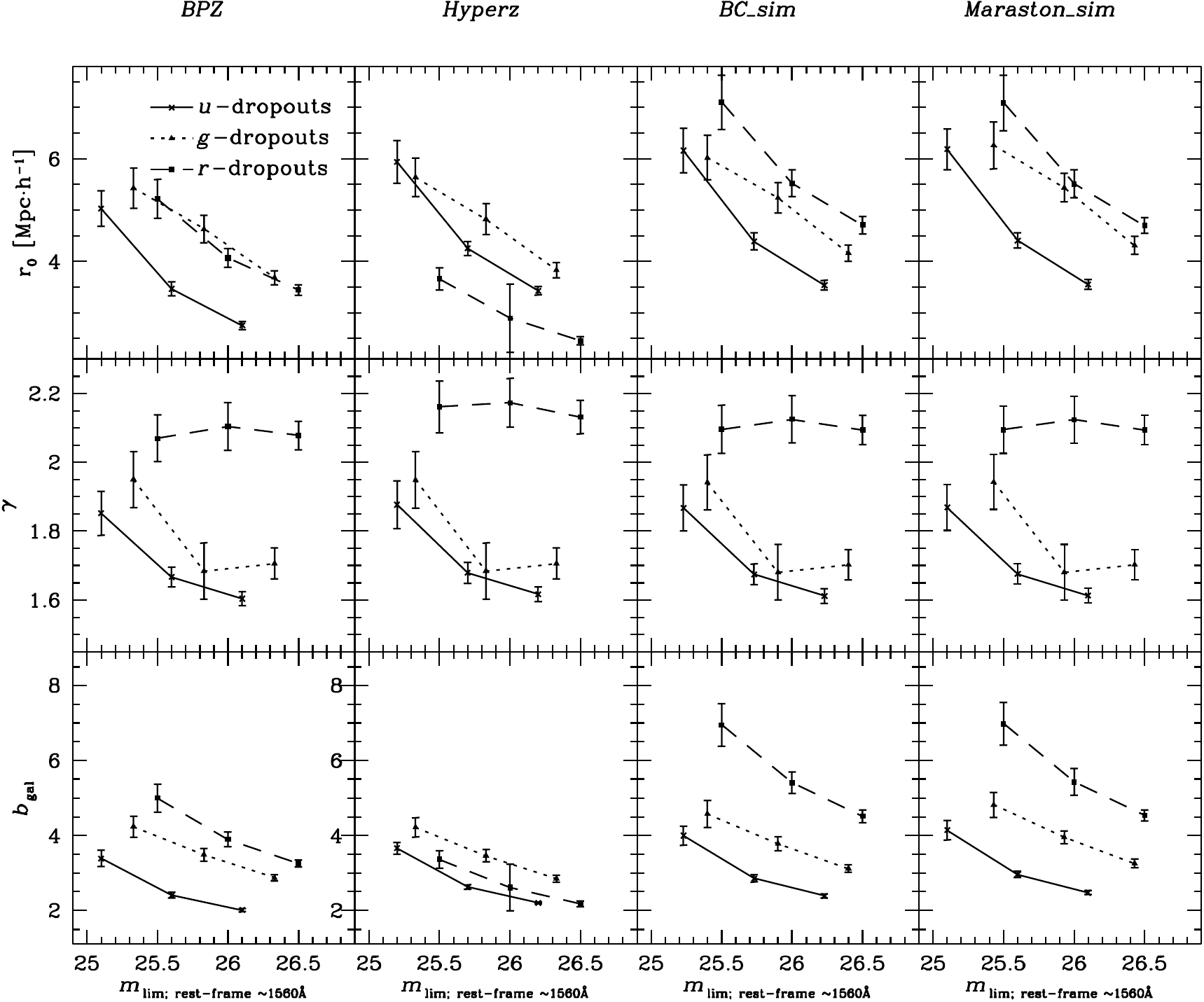}}
\caption{Dependence of the correlation length, the correlation
  function slope, and the galaxy bias parameter on redshift,
  UV-luminosity, and the assumed redshift distribution. The
  \emph{solid} lines and the crosses correspond to the $u$-dropouts,
  the \emph{dotted} lines and the triangles to the $g$-dropouts, and
  the \emph{dashed} lines and the squares to the $r$-dropouts,
  respectively. We scale the $r$- and $i$-band limiting magnitudes of
  the $u$- and $g$-dropouts to the $z$-band limits of the $r$-dropouts
  as described in the text.}
\label{fig:r_0}
\end{figure*}

The results are summarised in
Tables~\ref{tab:r_0_z_mag_BPZ}-\ref{tab:r_0_z_mag_Masim} and the
dependence of different measured quantities on redshift, magnitude,
and the assumed redshift distribution is visualised in
Fig.~\ref{fig:r_0}. The errors for $r_0$ and $b_{\rm gal}$ in
Fig.~\ref{fig:r_0} are estimated in Monte Carlo realisations from the
errors of the power-law fit. Therein, we assume uncorrelated Gaussian
errors of the amplitude and the slope of the power-law. In the lower
S/N domain the error introduced by the binning of the data becomes
dominant over the fitting error. We add this contribution in
quadrature, which is estimated from the scatter for many different
binnings.

\begin{table*}
\begin{minipage}[t]{\columnwidth}
\caption{Clustering measurements of the three dropout samples for
  different flux limits using the \emph{BPZ} redshift distributions.}
\label{tab:r_0_z_mag_BPZ}
\renewcommand{\footnoterule}{}
\begin{tabular}{l|ccccc}
\hline
\hline
&\multicolumn{5}{c}{$u$-dropouts; $z_{\rm mean}=3.28$}\\
\hline
$r_{\mathrm lim}$ & $n_{\rm g}$   & $\gamma$ & $r_0$ & $b_{\rm gal}$ & IC \\
                & [$h^3{\rm Mpc}^{-3}$] &          & [$h^{-1}$Mpc] & & \\
\hline
24.5 & $5.45\times 10^{-4}\pm5.55\times 10^{-5}$ & $1.85\pm0.06$ & $5.03\pm0.35$ & $3.39\pm0.22$ & 0.0056 \\
25.0 & $1.58\times 10^{-3}\pm1.59\times 10^{-4}$ & $1.67\pm0.03$ & $3.47\pm0.13$ & $2.41\pm0.08$ & 0.0028 \\
25.5 & $3.43\times 10^{-3}\pm3.44\times 10^{-4}$ & $1.60\pm0.02$ & $2.76\pm0.08$ & $2.01\pm0.05$ & 0.0020 \\
\hline
\hline
$i_{\mathrm lim}$&\multicolumn{5}{c}{$g$-dropouts; $z_{\rm mean}=3.87$}\\
\hline
25.0 & $5.20\times 10^{-4}\pm5.28\times 10^{-5}$ & $1.95\pm0.08$ & $5.43\pm0.39$ & $4.24\pm0.28$ & 0.0021 \\
25.5 & $1.32\times 10^{-3}\pm1.33\times 10^{-4}$ & $1.68\pm0.08$ & $4.64\pm0.27$ & $3.49\pm0.17$ & 0.0014 \\
26.0 & $2.67\times 10^{-3}\pm2.68\times 10^{-4}$ & $1.71\pm0.05$ & $3.69\pm0.13$ & $2.87\pm0.08$ & 0.0010 \\
\hline
\hline
$z_{\mathrm lim}$&\multicolumn{5}{c}{$r$-dropouts; $z_{\rm mean}=4.74$}\\
\hline
25.5 & $4.96\times 10^{-4}\pm5.11\times 10^{-5}$ & $2.07\pm0.07$ & $5.22\pm0.38$ & $5.00\pm0.37$ & 0.0142 \\
26.0 & $1.18\times 10^{-3}\pm1.19\times 10^{-4}$ & $2.10\pm0.07$ & $4.07\pm0.18$ & $3.90\pm0.19$ & 0.0086 \\
26.5 & $2.08\times 10^{-3}\pm2.09\times 10^{-4}$ & $2.08\pm0.04$ & $3.45\pm0.10$ & $3.25\pm0.09$ & 0.0060 \\
\hline
\end{tabular}\\
\end{minipage}
\end{table*}

\begin{table*}
\begin{minipage}[t]{\columnwidth}
\caption{Same as Table~\ref{tab:r_0_z_mag_BPZ} but for the
  \emph{Hyperz} redshift distributions.}
\label{tab:r_0_z_mag_BC}
\renewcommand{\footnoterule}{}
\begin{tabular}{l|ccccc}
\hline
\hline
&\multicolumn{5}{c}{$u$-dropouts; $z_{\rm mean}=2.93$}\\
\hline
$r_{\mathrm lim}$ & $n_{\rm g}$   & $\gamma$ & $r_0$ & $b_{\rm gal}$ & IC \\
                & [$h^3{\rm Mpc}^{-3}$] &          & [$h^{-1}$Mpc] & & \\
\hline
24.5 & $5.19\times 10^{-4}\pm5.29\times 10^{-5}$ & $1.88\pm0.07$ & $5.94\pm0.42$ & $3.66\pm0.15$ & 0.0025 \\
25.0 & $1.50\times 10^{-3}\pm1.51\times 10^{-4}$ & $1.68\pm0.03$ & $4.25\pm0.13$ & $2.63\pm0.06$ & 0.0013 \\
25.5 & $3.27\times 10^{-3}\pm3.28\times 10^{-4}$ & $1.62\pm0.02$ & $3.43\pm0.08$ & $2.20\pm0.02$ & 0.0009 \\
\hline
\hline
$i_{\mathrm lim}$&\multicolumn{5}{c}{$g$-dropouts; $z_{\rm mean}=3.67$}\\
\hline
25.0 & $5.05\times 10^{-4}\pm5.13\times 10^{-5}$ & $1.95\pm0.08$ & $5.64\pm0.38$ & $4.22\pm0.26$ & 0.0021 \\
25.5 & $1.29\times 10^{-3}\pm1.29\times 10^{-4}$ & $1.68\pm0.08$ & $4.83\pm0.30$ & $3.46\pm0.17$ & 0.0014 \\
26.0 & $2.60\times 10^{-3}\pm2.61\times 10^{-4}$ & $1.71\pm0.05$ & $3.83\pm0.15$ & $2.85\pm0.10$ & 0.0010 \\
\hline
\hline
$z_{\mathrm lim}$&\multicolumn{5}{c}{$r$-dropouts; $z_{\rm mean}=4.47$}\\
\hline
25.5 & $9.80\times 10^{-4}\pm1.01\times 10^{-4}$ & $2.16\pm0.08$ & $3.66\pm0.22$ & $3.36\pm0.23$ & 0.0031 \\
26.0 & $2.33\times 10^{-3}\pm2.36\times 10^{-4}$ & $2.17\pm0.07$ & $2.90\pm0.67$ & $2.61\pm0.62$ & 0.0018 \\
26.5 & $4.10\times 10^{-3}\pm4.13\times 10^{-4}$ & $2.13\pm0.05$ & $2.46\pm0.07$ & $2.18\pm0.08$ & 0.0013 \\
\hline
\end{tabular}\\
\end{minipage}
\end{table*}

\begin{table*}
\begin{minipage}[t]{\columnwidth}
\caption{Same as Table~\ref{tab:r_0_z_mag_BPZ} but for the
  \emph{BC\_sim} redshift distributions including halo model
  estimates.}
\label{tab:r_0_z_mag_BCsim}
\renewcommand{\footnoterule}{}
\begin{tabular}{l|ccccccc}
\hline
\hline
&\multicolumn{7}{c}{$u$-dropouts; $z_{\rm mean}=3.16$}\\
\hline
$r_{\mathrm lim}$ & $n_{\rm g}$   & $\gamma$ & $r_0$ & $b_{\rm gal}$ & IC & $\log\left<M_{\mathrm{halo}}\right>$ & $\left<N_g\right>$ \\
                & [$h^3{\rm Mpc}^{-3}$] &          & [$h^{-1}$Mpc] & & & [$h^{-1}M_\odot$] & \\
\hline
24.5 & $4.17\times10^{-4}\pm4.24\times10^{-5}$ & $1.87\pm0.07$ & $6.16\pm0.43$ & $4.00\pm0.26$ & 0.0036 & $12.68_{-0.37}^{+0.20}$ & $0.32\pm0.27$ \\
25.0 & $1.21\times10^{-3}\pm1.21\times10^{-4}$ & $1.68\pm0.03$ & $4.39\pm0.17$ & $2.86\pm0.09$ & 0.0018 & $12.26_{-0.33}^{+0.19}$ & $0.70\pm1.02$ \\
25.5 & $2.62\times10^{-3}\pm2.63\times10^{-4}$ & $1.61\pm0.02$ & $3.54\pm0.10$ & $2.39\pm0.05$ & 0.0013 & $12.06_{-0.19}^{+0.13}$ & $0.84\pm0.66$ \\
\hline
\hline
$i_{\mathrm lim}$&\multicolumn{7}{c}{$g$-dropouts; $z_{\rm mean}=3.76$}\\
\hline
25.0 & $3.41\times10^{-4}\pm3.46\times10^{-5}$ & $1.94\pm0.08$ & $6.02\pm0.43$ & $4.57\pm0.36$ & 0.0028 & $12.39_{-0.13}^{+0.10}$ & $0.45\pm0.33$ \\
25.5 & $8.68\times10^{-4}\pm8.73\times10^{-5}$ & $1.68\pm0.08$ & $5.24\pm0.30$ & $3.78\pm0.19$ & 0.0019 & $12.17_{-0.07}^{+0.06}$ & $0.65\pm0.57$ \\
26.0 & $1.76\times10^{-3}\pm1.76\times10^{-4}$ & $1.70\pm0.04$ & $4.16\pm0.16$ & $3.12\pm0.10$ & 0.0013 & $12.08_{-0.06}^{+0.05}$ & $0.48\pm0.13$ \\
\hline
\hline
$z_{\mathrm lim}$&\multicolumn{7}{c}{$r$-dropouts; $z_{\rm mean}=4.73$}\\
\hline
25.5 & $2.27\times10^{-4}\pm2.34\times10^{-5}$ & $2.10\pm0.07$ & $7.10\pm0.53$ & $6.95\pm0.57$ & 0.0108 & $12.26_{-0.17}^{+0.12}$ & $0.30\pm0.18$ \\
26.0 & $5.40\times10^{-4}\pm5.46\times10^{-5}$ & $2.13\pm0.07$ & $5.53\pm0.26$ & $5.41\pm0.29$ & 0.0066 & $12.08_{-0.25}^{+0.16}$ & $0.39\pm0.26$ \\
26.5 & $9.49\times10^{-4}\pm9.56\times10^{-5}$ & $2.09\pm0.04$ & $4.71\pm0.17$ & $4.52\pm0.17$ & 0.0046 & $12.00_{-0.19}^{+0.13}$ & $0.16\pm0.20$ \\
\hline
\end{tabular}\\
\end{minipage}
\end{table*}

\begin{table*}
\begin{minipage}[t]{\columnwidth}
\caption{Same as Table~\ref{tab:r_0_z_mag_BPZ} but for the
  \emph{Maraston\_sim} redshift distributions.}
\label{tab:r_0_z_mag_Masim}
\renewcommand{\footnoterule}{}
\begin{tabular}{l|ccccc}
\hline
\hline
&\multicolumn{5}{c}{$u$-dropouts; $z_{\rm mean}=3.30$}\\
\hline
$r_{\mathrm lim}$ & $n_{\rm g}$   & $\gamma$ & $r_0$ & $b_{\rm gal}$ & IC \\
                & [$h^3{\rm Mpc}^{-3}$] &          & [$h^{-1}$Mpc] & & \\
\hline
24.5 & $3.86\times 10^{-4}\pm3.93\times 10^{-5}$ & $1.87\pm0.07$ & $6.19\pm0.40$ & $4.14\pm0.26$ & 0.0035 \\
25.0 & $1.12\times 10^{-3}\pm1.12\times 10^{-4}$ & $1.68\pm0.03$ & $4.41\pm0.15$ & $2.96\pm0.09$ & 0.0018 \\
25.5 & $2.43\times 10^{-3}\pm2.43\times 10^{-4}$ & $1.61\pm0.02$ & $3.55\pm0.10$ & $2.48\pm0.05$ & 0.0012 \\
\hline
\hline
$i_{\mathrm lim}$&\multicolumn{5}{c}{$g$-dropouts; $z_{\rm mean}=3.83$}\\
\hline
25.0 & $3.10\times 10^{-4}\pm3.15\times 10^{-5}$ & $1.94\pm0.08$ & $6.26\pm0.46$ & $4.82\pm0.33$ & 0.0028 \\
25.5 & $7.89\times 10^{-4}\pm7.94\times 10^{-5}$ & $1.68\pm0.08$ & $5.44\pm0.28$ & $3.95\pm0.17$ & 0.0019 \\
26.0 & $1.60\times 10^{-3}\pm1.60\times 10^{-4}$ & $1.70\pm0.04$ & $4.32\pm0.17$ & $3.26\pm0.11$ & 0.0013 \\
\hline
\hline
$z_{\mathrm lim}$&\multicolumn{5}{c}{$r$-dropouts; $z_{\rm mean}=4.77$}\\
\hline
25.5 & $2.12\times 10^{-4}\pm2.19\times 10^{-5}$ & $2.09\pm0.07$ & $7.09\pm0.54$ & $6.98\pm0.56$ & 0.0110 \\
26.0 & $5.05\times 10^{-4}\pm5.11\times 10^{-5}$ & $2.12\pm0.07$ & $5.51\pm0.27$ & $5.43\pm0.36$ & 0.0067 \\
26.5 & $8.88\times 10^{-4}\pm8.94\times 10^{-5}$ & $2.09\pm0.04$ & $4.70\pm0.16$ & $4.54\pm0.16$ & 0.0046 \\
\hline
\end{tabular}\\
\end{minipage}
\end{table*}

There is clear evidence for clustering segregation with rest-frame UV
luminosity for all three dropout samples regardless of the assumed
redshift distribution. Brighter samples cluster more strongly than
fainter ones as reflected by their correlation lengths and bias
values.

The observation that the correlation function slope on intermediate
and large scales gets shallower for fainter galaxies as reported by
\cite{2005ApJ...635L.117O}, \cite{2006ApJ...637..631K}, and
\cite{2007A&A...462..865H} is also detected in our $u$- and
$g$-dropout samples. Furthermore, we also detect a similar
levelling-off of the slope for the two faintest bins as reported by
\cite{2007A&A...462..865H} for a $U$-dropout sample. The flattening of
the correlation function for fainter magnitudes is not observed for
the $r$-dropouts. They show a constant, rather steep slope of
$\gamma\approx2.1$ for all magnitude limits.

The measurement of correlation lengths is strongly dependent on the
assumed redshift distributions which have not been spectroscopically
verified.  However, the $u$-dropout samples show smaller correlation
lengths than the higher redshift samples for the \emph{BPZ}, the
\emph{BC\_sim}, and the \emph{Maraston\_sim} redshift distributions,
indicating some evolution may be occurring.

A much clearer redshift dependence is observed for the values of the
large-scale galaxy bias again for the \emph{BPZ}, the \emph{BC\_sim},
and the \emph{Maraston\_sim} redshift distributions. From $z\sim5$ to
$z\sim3$ the galaxy bias decreases by a factor of two. This more
pronounced trend in comparison to the correlation lengths can easily
be explained by the ongoing structure formation in the early
universe. DM halos with roughly the same comoving correlation length
are more biased tracers of the underlying dark matter field at high
redshifts than at lower redshifts. This is due to the fact that the
amplitudes of the overall DM density field grow with time. Thus, the
scale where the correlation function takes a value of one (the
correlation lengths) increases over time.

In \cite{2007A&A...462..865H} we found clustering segregation with UV
luminosity as well but the correlation lengths of that $U$-dropout
sample were consistent with the ones from the $B$-dropout sample from
\cite{2005ApJ...635L.117O} for the same luminosity, showing no
evolution with redshift. Besides the much larger statistical power of
the present survey, it should be noted that several systematic effects
can influence such a comparison of different datasets. For example, it
is not clear whether intrinsically very similar populations were
studied in \cite{2005ApJ...635L.117O} and \cite{2007A&A...462..865H}
because of the different filter sets, the different selection
criteria, and the different depths of the data. Furthermore, masking
can introduce biases in the clustering signal that are hard to
control. Most importantly (see above), the redshift distributions were
obtained in very different ways with \cite{2007A&A...462..865H}
relying on \emph{Hyperz} photo-$z$'s and \cite{2005ApJ...635L.117O} on
simulations. Indeed, the non-evolution of the correlation length with
redshift from $z\sim4$ to $z\sim3$ is seen also here in the
\emph{Hyperz} panel in Fig.~\ref{fig:r_0}. It might well be spurious
due to the problems mentioned in Sect.~\ref{sec:hyperz}.

In the present study we carry out a coherent analysis on one
dataset. The selection criteria are taken from the same set of
simulations and the masking is identical for all three
samples. However, without spectroscopic redshift distributions we
cannot decide whether the evolutionary trend observed with the
\emph{BPZ}, the \emph{BC\_sim}, and the \emph{Maraston\_sim} redshift
distributions or the non-evolution observed for the \emph{Hyperz}
redshift distribution is real. The simulated redshift distributions
have the advantage that they are not affected by a number of
systematic errors inherent to the observations (except for the
reliance on the observed $i$-band number counts) or spurious effects
introduced by priors in the photo-$z$ codes. Furthermore, the absolute
results from the \emph{BC\_sim} case and the \emph{Maraston\_sim} case
agree very well, although different template sets were used. One
disadvantage of the simulations is that both rely on assumptions about
the mixture of templates and the fractions of high- to low-$z$ objects
which are taken from external data (the HDF-N and the CFRS). We tend
to trust the results based on the simulated redshift distributions
more, but note that a decision about redshift evolution or
non-evolution cannot be taken without additional massive spectroscopic
support.

The same is true for the influence of contamination or incompleteness
on the clustering measurements. Without a deep spectroscopic survey we
cannot account for such effects because we do not know how these
possible low-$z$ contaminants cluster or which galaxies at high-$z$
are missed by the Lyman-break selection. This is a fundamental
limitation of a photometric survey. However, our simulations suggest
-- and the colour selection criteria and magnitude cuts were chosen in
such a way -- that contamination and incompleteness are kept
low. Thus, we still think it is reasonable to assume that the
clustering measurements are not seriously affected by either of the
two effects.

\subsection{Halo model}
As detected for $B$-dropouts
\citep{2005ApJ...635L.117O,2006ApJ...642...63L} and $U$-dropouts
\citep{2007A&A...462..865H} before, we see a significant deviation of
the angular correlation function on small scales from the power-law
behaviour on large scales for all three dropout samples. Especially,
for the first time we detect this one-halo term contribution to the
correlation function in our $r$-dropout sample at redshifts
considerably larger than $z=4$. This allows us to fit a halo model to
our data and estimate mean DM halo masses and occupation numbers
(i.e. the mean number of galaxies hosted by a halo) for the different
subsamples.

We apply the halo model with the same parameters as
\citet{2004MNRAS.347..813H} which was used in several high-$z$
clustering studies before \citep[e.g. in ][]{2005ApJ...635L.117O,
  2007A&A...462..865H}. More advanced halo models are available
\citep[see e.g. ][]{2008arXiv0808.1727L} but we decided to stick to
the more simple one in order to keep our results comparable to older
measurements. We only show the case for the \emph{BC\_sim} redshift
distributions since the dependence of halo masses on the assumed
redshift distribution is similar to the dependence of the correlation
lengths and the galaxy bias values (see
Sect.~\ref{sec:clustering_results}).

We fit for the mean galaxy density, $n_{\rm g}$, and the angular
correlation function simultaneously. The mean galaxy density (also
tabulated in Table~\ref{tab:r_0_z_mag_BCsim}) is estimated from the
observed number of objects divided by the survey volume, the latter
being estimated from the unmasked area and the redshift
distribution. The error of the mean galaxy density is estimated from
Poissonian variance of the galaxy numbers and from assuming a 10\%
error on the survey volume. We choose the angular fitting range for
the halo models to $0\farcm02<\theta<10'$.

\begin{figure}
\centering
\resizebox{\hsize}{!}{\includegraphics{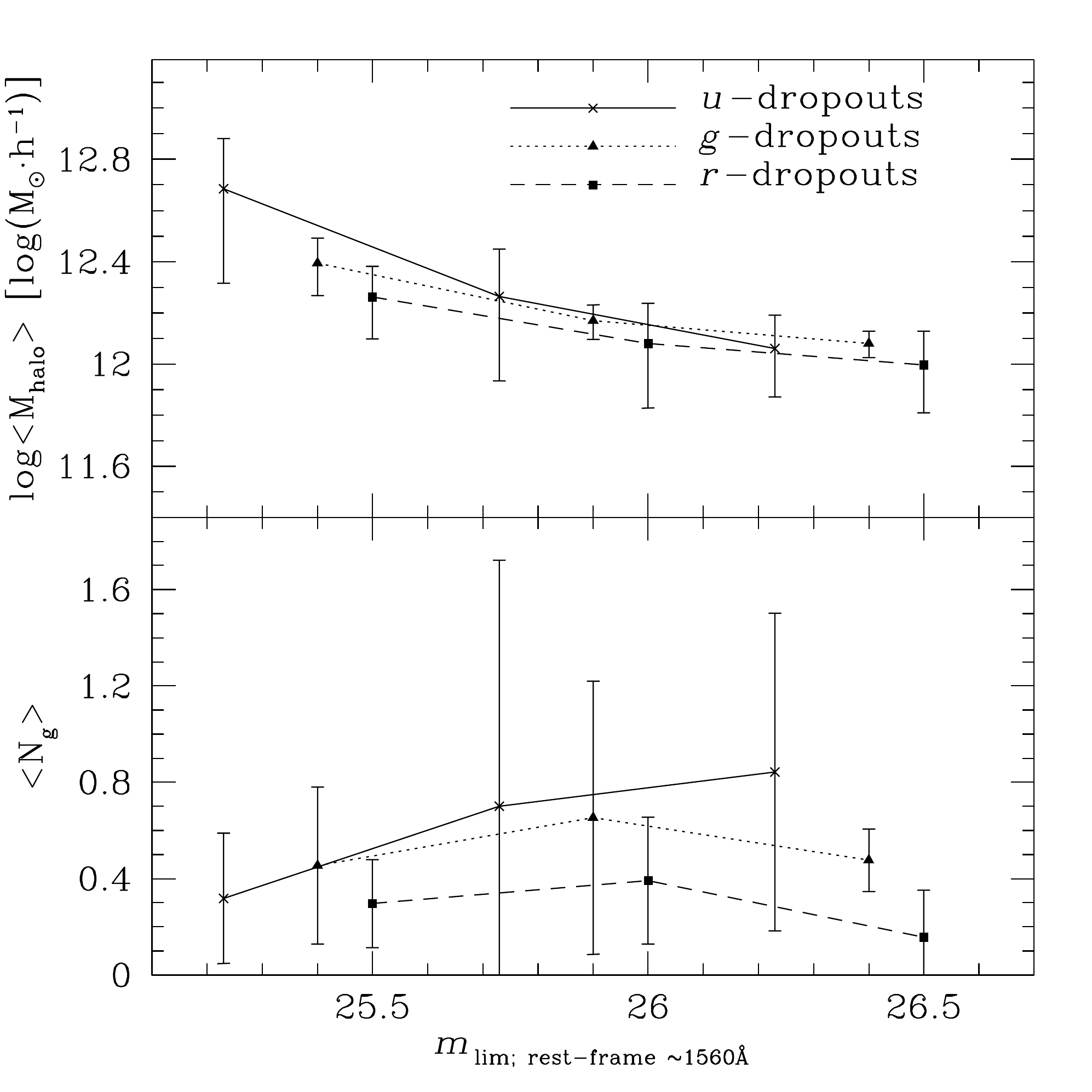}}
\caption{Dependence of the mean halo mass and the mean halo occupation
  number on redshift, UV-luminosity for the \emph{BC\_sim} redshift
  distribution. Symbols and scalings same as Fig.~\ref{fig:r_0}}
\label{fig:halo}
\end{figure}

The inferred mean halo masses and mean halo occupation numbers for the
different flux-limited subsamples are also listed in
Table~\ref{tab:r_0_z_mag_BCsim} and displayed in
Fig.~\ref{fig:halo}. Errors are again estimated from Monte-Carlo
realisations assuming uncorrelated Gaussian errors on the halo model
parameters $M_{\rm min}$, $M_1$, and $\alpha$, and we add the errors
derived from the fluctuations introduced by the binning of the data.

The halo masses for the three redshift samples are within errors
nearly consistent with each other at fixed luminosity. While we see
again a trend in the halo masses with luminosity in all three samples,
a significant redshift dependence is not observed. Our data seem to
favour the observation also made before by \cite{2007A&A...462..865H}
that lower redshift LBGs are hosted by more massive halos than higher
redshift LBGs of the same rest-frame UV-luminosity. If this trend is
real, the star-formation rate per unit mass decreases with redshift --
at least in LBGs. This result is, however, highly dependent on the
assumed redshift distribution. Taking e.g. the \emph{BPZ} redshift
distribution would further wash-out this already weak redshift
evolution.

The mean halo occupation numbers are not very well constrained by the
fit to the data but the trend that $\left<N_g\right><1$ for LBGs (as
observed before by
\citeauthor{2005ApJ...635L.117O}~\citeyear{2005ApJ...635L.117O} and
\citeauthor{2007A&A...462..865H}~\citeyear{2007A&A...462..865H}) is
confirmed. The large errors seem to originate from some tension
between the observed angular correlation functions and the observed
number densities.\footnote{See also \cite{2008ApJ...685L...1Q} for a
  more severe case of such a discrepancy between the shape of the
  correlation function and the number density observed with distant
  red galaxies.} This leads to only weak constrains on the $M_1$
parameter which is important to calculate $\left<N_g\right>$, whereas
it does not influence the estimate of the $\left<M_{\rm
  halo}\right>$. A more elaborate method of calculating the errors on
$\left<N_g\right>$ dropping the assumptions of Gaussianity and
uncorrelated errors of the halo model parameters could potentially
decrease uncertainties here.

Together with the observations of the strong one-halo term these
occupation numbers mean that most halos of the mean mass reported here
are devoid of LBGs and only a small number of them host a galaxy or a
small group.

\section{Conclusions}
\label{sec:conclusions}
We carried out a precision analysis of LBG clustering at different
redshifts from extremely deep optical imaging data in the CFHTLS-Deep
fields. The LBG samples presented here are by far the largest studied
so far. We detect a clear evolution of the large-scale galaxy bias and
a tentative evolution of the correlation lengths with redshift for
three of the four redshift distributions considered here. Higher-$z$
LBGs cluster more strongly than lower-$z$ ones. In agreement with
former studies, we observe a strong trend with luminosity with
brighter LBGs showing larger correlation lengths and galaxy bias values
than fainter ones. Since the samples were selected from one dataset
and the whole clustering analysis was done in a coherent way we can be
confident that the relative comparisons are more accurate than
comparisons carried out before which involved data from different
sources. Also the absolute values of the correlation lengths, bias
values, and slopes of the correlation functions are in good agreement
to former studies considering the systematic difficulties arising from
a redshift distribution based on simulations and the possible
contribution of contaminants to the clustering signal \citep[see also
][ for a detailed discussion of this aspect]{2008MNRAS.385..493S}.

The common picture of LBGs forming in massive halos of
$M\ga10^{12}h^{-1}M_{\odot}$ is supported by our halo model
analysis. Predictions from models of galaxy formation can be compared
to ever more detailed observational results in this way. By detecting
a significant one-halo term in the correlation function of
$r$-dropouts for the first time, we have observational evidence for
multiple galaxies hosted by single DM halos at very high redshifts.

In order to take this method to even higher precision and constrain
models of galaxy formation and evolution even better the systematic
errors must be better understood and controlled. A spectroscopic
redshift distribution of representative subsamples is inevitable for
reaching this goal. On the theoretical side one must think about more
realistic models for galaxy clustering since this particular halo
model used here is not a very good fit to the data, leaving
considerable residuals. \cite{2008arXiv0808.1727L} introduce
interesting extensions to the halo model used here. We plan to
incorporate information on the luminosity functions of the LBGs in a
future study to learn more about the relationship between the star
formation and the DM structure at high redshift. However, here we
stuck to the simpler model to keep results comparable to older
studies.

\begin{acknowledgements}
We would like to thank Ryan Quadri for interesting discussions about
galaxy clustering measurements that helped to improve the paper.  Many
thanks also to Claudia Maraston and Janine Pforr for help with their
template set. We are grateful to the CFHTLS survey team for conducting
the observations and the TERAPIX team for developing software used in
this study. We acknowledge use of the Canadian Astronomy Data Centre
operated by the Dominion Astrophysical Observatory for the National
Research Council of Canada's Herzberg Institute of Astrophysics.  This
work was supported by the DFG priority program SPP-1177 "Witnesses of
Cosmic History: Formation and evolution of black holes, galaxies and
their environment" (project ID ER327/2-2), the German Ministry for
Science and Education (BMBF) through DESY under the project 05AV5PDA/3
and the TR33 "The Dark Universe". HH and PS were supported by the
European DUEL RTN, project MRTN-CT-2006-036133. LVW was supported by
the Canadian Foundation for Innovation, NSERC and CIfAR.
\end{acknowledgements}

\bibliographystyle{aa}

\bibliography{1042}

\begin{thebibliography}{47}
\expandafter\ifx\csname natexlab\endcsname\relax\def\natexlab#1{#1}\fi

\bibitem[{{Adelberger} {et~al.}(2005){Adelberger}, {Steidel}, {Pettini},
  {Shapley}, {Reddy}, \& {Erb}}]{2005ApJ...619..697A}
{Adelberger}, K.~L., {Steidel}, C.~C., {Pettini}, M., {et~al.} 2005, \apj, 619,
  697

\bibitem[{{Ben{\'{\i}}tez}(2000)}]{2000ApJ...536..571B}
{Ben{\'{\i}}tez}, N. 2000, \apj, 536, 571

\bibitem[{{Bertin}(2003)}]{2003SWarp}
{Bertin}, E. 2003, SWarp user's guide, \url{http://terapix.iap.fr}

\bibitem[{{Bertin} \& {Arnouts}(1996)}]{1996A&AS..117..393B}
{Bertin}, E. \& {Arnouts}, S. 1996, \aaps, 117, 393

\bibitem[{{Bolzonella} {et~al.}(2000){Bolzonella}, {Miralles}, \& {Pell{\'
  o}}}]{2000A&A...363..476B}
{Bolzonella}, M., {Miralles}, J.-M., \& {Pell{\' o}}, R. 2000, \aap, 363, 476

\bibitem[{{Bruzual } \& {Charlot}(1993)}]{1993ApJ...405..538B}
{Bruzual }, A.~G. \& {Charlot}, S. 1993, \apj, 405, 538

\bibitem[{{Budav{\'a}ri} {et~al.}(1999){Budav{\'a}ri}, {Szalay}, {Connolly},
  {Csabai}, {Dickinson}, \& {The Hdf/Nicmos Team}}]{1999ASPC..191...19B}
{Budav{\'a}ri}, T., {Szalay}, A.~S., {Connolly}, A.~J., {et~al.} 1999, in
  Astronomical Society of the Pacific Conference Series, Vol. 191, Photometric
  Redshifts and the Detection of High Redshift Galaxies, ed. R.~{Weymann},
  L.~{Storrie-Lombardi}, M.~{Sawicki}, \& R.~{Brunner}, 19

\bibitem[{{Capak}(2004)}]{2004_Capak_PhDT}
{Capak}, P.~L. 2004, PhD thesis, AA(UNIVERSITY OF HAWAI'I)

\bibitem[{{Coleman} {et~al.}(1980){Coleman}, {Wu}, \&
  {Weedman}}]{1980ApJS...43..393C}
{Coleman}, G.~D., {Wu}, C.-C., \& {Weedman}, D.~W. 1980, \apjs, 43, 393

\bibitem[{{Cooray} \& {Sheth}(2002)}]{2002PhR...372....1C}
{Cooray}, A. \& {Sheth}, R. 2002, \physrep, 372, 1

\bibitem[{{Davis} {et~al.}(2007){Davis}, {Guhathakurta}, {Konidaris}, {Newman},
  {Ashby}, {Biggs}, {Barmby}, {Bundy}, {Chapman}, {Coil}, {Conselice},
  {Cooper}, {Croton}, {Eisenhardt}, {Ellis}, {Faber}, {Fang}, {Fazio},
  {Georgakakis}, {Gerke}, {Goss}, {Gwyn}, {Harker}, {Hopkins}, {Huang},
  {Ivison}, {Kassin}, {Kirby}, {Koekemoer}, {Koo}, {Laird}, {Le Floc'h}, {Lin},
  {Lotz}, {Marshall}, {Martin}, {Metevier}, {Moustakas}, {Nandra}, {Noeske},
  {Papovich}, {Phillips}, {Rich}, {Rieke}, {Rigopoulou}, {Salim},
  {Schiminovich}, {Simard}, {Smail}, {Small}, {Weiner}, {Willmer}, {Willner},
  {Wilson}, {Wright}, \& {Yan}}]{2007ApJ...660L...1D}
{Davis}, M., {Guhathakurta}, P., {Konidaris}, N.~P., {et~al.} 2007, \apjl, 660,
  L1

\bibitem[{{Erben} {et~al.}(2009){Erben}, {Hildebrandt}, {Lerchster}, {Hudelot},
  {Benjamin}, {van Waerbeke}, {Schrabback}, {Brimioulle}, {Cordes}, {Dietrich},
  {Holhjem}, {Schirmer}, \& {Schneider}}]{2009A&A...493.1197E}
{Erben}, T., {Hildebrandt}, H., {Lerchster}, M., {et~al.} 2009, \aap, 493, 1197

\bibitem[{{Erben} {et~al.}(2005){Erben}, {Schirmer}, {Dietrich}, {Cordes},
  {Haberzettl}, {Hetterscheidt}, {Hildebrandt}, {Schmithuesen}, {Schneider},
  {Simon}, {Deul}, {Hook}, {Kaiser}, {Radovich}, {Benoist}, {Nonino}, {Olsen},
  {Prandoni}, {Wichmann}, {Zaggia}, {Bomans}, {Dettmar}, \&
  {Miralles}}]{2005AN....326..432E}
{Erben}, T., {Schirmer}, M., {Dietrich}, J.~P., {et~al.} 2005, Astronomische
  Nachrichten, 326, 432

\bibitem[{{Fern{\'a}ndez-Soto} {et~al.}(1999){Fern{\'a}ndez-Soto}, {Lanzetta},
  \& {Yahil}}]{1999ApJ...513...34F}
{Fern{\'a}ndez-Soto}, A., {Lanzetta}, K.~M., \& {Yahil}, A. 1999, \apj, 513, 34

\bibitem[{{Giavalisco}(2002)}]{2002ARA&A..40..579G}
{Giavalisco}, M. 2002, \araa, 40, 579

\bibitem[{{Girardi} {et~al.}(2005){Girardi}, {Groenewegen}, {Hatziminaoglou},
  \& {da Costa}}]{2005A&A...436..895G}
{Girardi}, L., {Groenewegen}, M.~A.~T., {Hatziminaoglou}, E., \& {da Costa}, L.
  2005, \aap, 436, 895

\bibitem[{{Hamana} {et~al.}(2004){Hamana}, {Ouchi}, {Shimasaku}, {Kayo}, \&
  {Suto}}]{2004MNRAS.347..813H}
{Hamana}, T., {Ouchi}, M., {Shimasaku}, K., {Kayo}, I., \& {Suto}, Y. 2004,
  \mnras, 347, 813

\bibitem[{{Hildebrandt} {et~al.}(2006){Hildebrandt}, {Erben}, {Dietrich},
  {Cordes}, {Haberzettl}, {Hetterscheidt}, {Schirmer}, {Schmithuesen},
  {Schneider}, {Simon}, \& {Trachternach}}]{2006A&A...452.1121H}
{Hildebrandt}, H., {Erben}, T., {Dietrich}, J.~P., {et~al.} 2006, \aap, 452,
  1121

\bibitem[{{Hildebrandt} {et~al.}(2007){Hildebrandt}, {Pielorz}, {Erben},
  {Schneider}, {Eifler}, {Simon}, \& {Dietrich}}]{2007A&A...462..865H}
{Hildebrandt}, H., {Pielorz}, J., {Erben}, T., {et~al.} 2007, \aap, 462, 865

\bibitem[{{Hildebrandt} {et~al.}(2008){Hildebrandt}, {Wolf}, \&
  {Ben{\'{\i}}tez}}]{2008A&A...480..703H}
{Hildebrandt}, H., {Wolf}, C., \& {Ben{\'{\i}}tez}, N. 2008, \aap, 480, 703

\bibitem[{{Kashikawa} {et~al.}(2006){Kashikawa}, {Yoshida}, {Shimasaku},
  {Nagashima}, {Yahagi}, {Ouchi}, {Matsuda}, {Malkan}, {Doi}, {Iye}, {Ajiki},
  {Akiyama}, {Ando}, {Aoki}, {Furusawa}, {Hayashino}, {Iwamuro}, {Karoji},
  {Kobayashi}, {Kodaira}, {Kodama}, {Komiyama}, {Miyazaki}, {Mizumoto},
  {Morokuma}, {Motohara}, {Murayama}, {Nagao}, {Nariai}, {Ohta}, {Okamura},
  {Sasaki}, {Sato}, {Sekiguchi}, {Shioya}, {Tamura}, {Taniguchi}, {Umemura},
  {Yamada}, \& {Yasuda}}]{2006ApJ...637..631K}
{Kashikawa}, N., {Yoshida}, M., {Shimasaku}, K., {et~al.} 2006, \apj, 637, 631

\bibitem[{{Kinney} {et~al.}(1996){Kinney}, {Calzetti}, {Bohlin}, {McQuade},
  {Storchi-Bergmann}, \& {Schmitt}}]{1996ApJ...467...38K}
{Kinney}, A.~L., {Calzetti}, D., {Bohlin}, R.~C., {et~al.} 1996, \apj, 467, 38

\bibitem[{{Landy} \& {Szalay}(1993)}]{1993ApJ...412...64L}
{Landy}, S.~D. \& {Szalay}, A.~S. 1993, \apj, 412, 64

\bibitem[{{Le F{\`e}vre} {et~al.}(2005){Le F{\`e}vre}, {Vettolani}, {Garilli},
  {Tresse}, {Bottini}, {Le Brun}, {Maccagni}, {Picat}, {Scaramella},
  {Scodeggio}, {Zanichelli}, {Adami}, {Arnaboldi}, {Arnouts}, {Bardelli},
  {Bolzonella}, {Cappi}, {Charlot}, {Ciliegi}, {Contini}, {Foucaud},
  {Franzetti}, {Gavignaud}, {Guzzo}, {Ilbert}, {Iovino}, {McCracken}, {Marano},
  {Marinoni}, {Mathez}, {Mazure}, {Meneux}, {Merighi}, {Paltani}, {Pell{\`o}},
  {Pollo}, {Pozzetti}, {Radovich}, {Zamorani}, {Zucca}, {Bondi}, {Bongiorno},
  {Busarello}, {Lamareille}, {Mellier}, {Merluzzi}, {Ripepi}, \&
  {Rizzo}}]{2005A&A...439..845L}
{Le F{\`e}vre}, O., {Vettolani}, G., {Garilli}, B., {et~al.} 2005, \aap, 439,
  845

\bibitem[{{Lee} {et~al.}(2008){Lee}, {Giavalisco}, {Conroy}, {Wechsler},
  {Ferguson}, {Somerville}, {Dickinson}, \& {Urry}}]{2008arXiv0808.1727L}
{Lee}, K., {Giavalisco}, M., {Conroy}, C., {et~al.} 2008, arXiv:0808.1727

\bibitem[{{Lee} {et~al.}(2006){Lee}, {Giavalisco}, {Gnedin}, {Somerville},
  {Ferguson}, {Dickinson}, \& {Ouchi}}]{2006ApJ...642...63L}
{Lee}, K.-S., {Giavalisco}, M., {Gnedin}, O.~Y., {et~al.} 2006, \apj, 642, 63

\bibitem[{{Lehnert} \& {Bremer}(2003)}]{2003ApJ...593..630L}
{Lehnert}, M.~D. \& {Bremer}, M. 2003, \apj, 593, 630

\bibitem[{{Lilly} {et~al.}(2007){Lilly}, {Le F{\`e}vre}, {Renzini}, {Zamorani},
  {Scodeggio}, {Contini}, {Carollo}, {Hasinger}, {Kneib}, {Iovino}, {Le Brun},
  {Maier}, {Mainieri}, {Mignoli}, {Silverman}, {Tasca}, {Bolzonella},
  {Bongiorno}, {Bottini}, {Capak}, {Caputi}, {Cimatti}, {Cucciati}, {Daddi},
  {Feldmann}, {Franzetti}, {Garilli}, {Guzzo}, {Ilbert}, {Kampczyk}, {Kovac},
  {Lamareille}, {Leauthaud}, {Borgne}, {McCracken}, {Marinoni}, {Pello},
  {Ricciardelli}, {Scarlata}, {Vergani}, {Sanders}, {Schinnerer}, {Scoville},
  {Taniguchi}, {Arnouts}, {Aussel}, {Bardelli}, {Brusa}, {Cappi}, {Ciliegi},
  {Finoguenov}, {Foucaud}, {Franceschini}, {Halliday}, {Impey}, {Knobel},
  {Koekemoer}, {Kurk}, {Maccagni}, {Maddox}, {Marano}, {Marconi}, {Meneux},
  {Mobasher}, {Moreau}, {Peacock}, {Porciani}, {Pozzetti}, {Scaramella},
  {Schiminovich}, {Shopbell}, {Smail}, {Thompson}, {Tresse}, {Vettolani},
  {Zanichelli}, \& {Zucca}}]{2007ApJS..172...70L}
{Lilly}, S.~J., {Le F{\`e}vre}, O., {Renzini}, A., {et~al.} 2007, \apjs, 172,
  70

\bibitem[{{Lilly} {et~al.}(1995){Lilly}, {Tresse}, {Hammer}, {Crampton}, \& {Le
  Fevre}}]{1995ApJ...455..108L}
{Lilly}, S.~J., {Tresse}, L., {Hammer}, F., {Crampton}, D., \& {Le Fevre}, O.
  1995, \apj, 455, 108

\bibitem[{{Limber}(1953)}]{1953ApJ...117..134L}
{Limber}, D.~N. 1953, \apj, 117, 134

\bibitem[{{Magnier} \& {Cuillandre}(2004)}]{2004PASP..116..449M}
{Magnier}, E.~A. \& {Cuillandre}, J.-C. 2004, \pasp, 116, 449

\bibitem[{{Maraston}(2005)}]{2005MNRAS.362..799M}
{Maraston}, C. 2005, \mnras, 362, 799

\bibitem[{{Maraston} {et~al.}(2006){Maraston}, {Daddi}, {Renzini}, {Cimatti},
  {Dickinson}, {Papovich}, {Pasquali}, \& {Pirzkal}}]{2006ApJ...652...85M}
{Maraston}, C., {Daddi}, E., {Renzini}, A., {et~al.} 2006, \apj, 652, 85

\bibitem[{{McCracken} {et~al.}(2003){McCracken}, {Radovich}, {Bertin},
  {Mellier}, {Dantel-Fort}, {Le F{\`e}vre}, {Cuillandre}, {Gwyn}, {Foucaud}, \&
  {Zamorani}}]{2003A&A...410...17M}
{McCracken}, H.~J., {Radovich}, M., {Bertin}, E., {et~al.} 2003, \aap, 410, 17

\bibitem[{{Mobasher} {et~al.}(2007){Mobasher}, {Capak}, {Scoville}, {Dahlen},
  {Salvato}, {Aussel}, {Thompson}, {Feldmann}, {Tasca}, {Lefevre}, {Lilly},
  {Carollo}, {Kartaltepe}, {McCracken}, {Mould}, {Renzini}, {Sanders},
  {Shopbell}, {Taniguchi}, {Ajiki}, {Shioya}, {Contini}, {Giavalisco},
  {Ilbert}, {Iovino}, {Le Brun}, {Mainieri}, {Mignoli}, \&
  {Scodeggio}}]{2007ApJS..172..117M}
{Mobasher}, B., {Capak}, P., {Scoville}, N.~Z., {et~al.} 2007, \apjs, 172, 117

\bibitem[{{Monet} {et~al.}(2003){Monet}, {Levine}, {Canzian}, {Ables}, {Bird},
  {Dahn}, {Guetter}, {Harris}, {Henden}, {Leggett}, {Levison}, {Luginbuhl},
  {Martini}, {Monet}, {Munn}, {Pier}, {Rhodes}, {Riepe}, {Sell}, {Stone},
  {Vrba}, {Walker}, {Westerhout}, {Brucato}, {Reid}, {Schoening}, {Hartley},
  {Read}, \& {Tritton}}]{2003AJ....125..984M}
{Monet}, D.~G., {Levine}, S.~E., {Canzian}, B., {et~al.} 2003, \aj, 125, 984

\bibitem[{{Ouchi} {et~al.}(2005){Ouchi}, {Hamana}, {Shimasaku}, {Yamada},
  {Akiyama}, {Kashikawa}, {Yoshida}, {Aoki}, {Iye}, {Saito}, {Sasaki},
  {Simpson}, \& {Yoshida}}]{2005ApJ...635L.117O}
{Ouchi}, M., {Hamana}, T., {Shimasaku}, K., {et~al.} 2005, \apjl, 635, L117

\bibitem[{{Peebles}(1980)}]{1980lssu.book.....P}
{Peebles}, P.~J.~E. 1980, {The large-scale structure of the universe}
  (Princeton University Press)

\bibitem[{{Quadri} {et~al.}(2008){Quadri}, {Williams}, {Lee}, {Franx}, {van
  Dokkum}, \& {Brammer}}]{2008ApJ...685L...1Q}
{Quadri}, R.~F., {Williams}, R.~J., {Lee}, K.-S., {et~al.} 2008, \apjl, 685, L1

\bibitem[{{Radovich}(2002)}]{2002ASTROMETRIX}
{Radovich}, M. 2002, ASTROMETRIX, \url{http://terapix.iap.fr}

\bibitem[{{Seljak}(2000)}]{2000MNRAS.318..203S}
{Seljak}, U. 2000, \mnras, 318, 203

\bibitem[{{Simon}(2007)}]{2007A&A...473..711S}
{Simon}, P. 2007, \aap, 473, 711

\bibitem[{{Stanway} {et~al.}(2008){Stanway}, {Bremer}, \&
  {Lehnert}}]{2008MNRAS.385..493S}
{Stanway}, E.~R., {Bremer}, M.~N., \& {Lehnert}, M.~D. 2008, \mnras, 385, 493

\bibitem[{{Steidel} {et~al.}(1999){Steidel}, {Adelberger}, {Giavalisco},
  {Dickinson}, \& {Pettini}}]{1999ApJ...519....1S}
{Steidel}, C.~C., {Adelberger}, K.~L., {Giavalisco}, M., {Dickinson}, M., \&
  {Pettini}, M. 1999, \apj, 519, 1

\bibitem[{{Steidel} {et~al.}(2003){Steidel}, {Adelberger}, {Shapley},
  {Pettini}, {Dickinson}, \& {Giavalisco}}]{2003ApJ...592..728S}
{Steidel}, C.~C., {Adelberger}, K.~L., {Shapley}, A.~E., {et~al.} 2003, \apj,
  592, 728

\bibitem[{{Steidel} {et~al.}(1996){Steidel}, {Giavalisco}, {Pettini},
  {Dickinson}, \& {Adelberger}}]{1996ApJ...462L..17S}
{Steidel}, C.~C., {Giavalisco}, M., {Pettini}, M., {Dickinson}, M., \&
  {Adelberger}, K.~L. 1996, \apjl, 462, L17

\bibitem[{{Vandame}(2001)}]{2001misk.conf..595V}
{Vandame}, B. 2001, in Mining the Sky, ed. A.~J. {Banday}, S.~{Zaroubi}, \&
  M.~{Bartelmann}, 595

\end{thebibliography}

\appendix

\section{Cross-correlation test of photo-$z$'s}
\label{app:cross}
\begin{figure*}
\centering
\includegraphics[width=17cm]{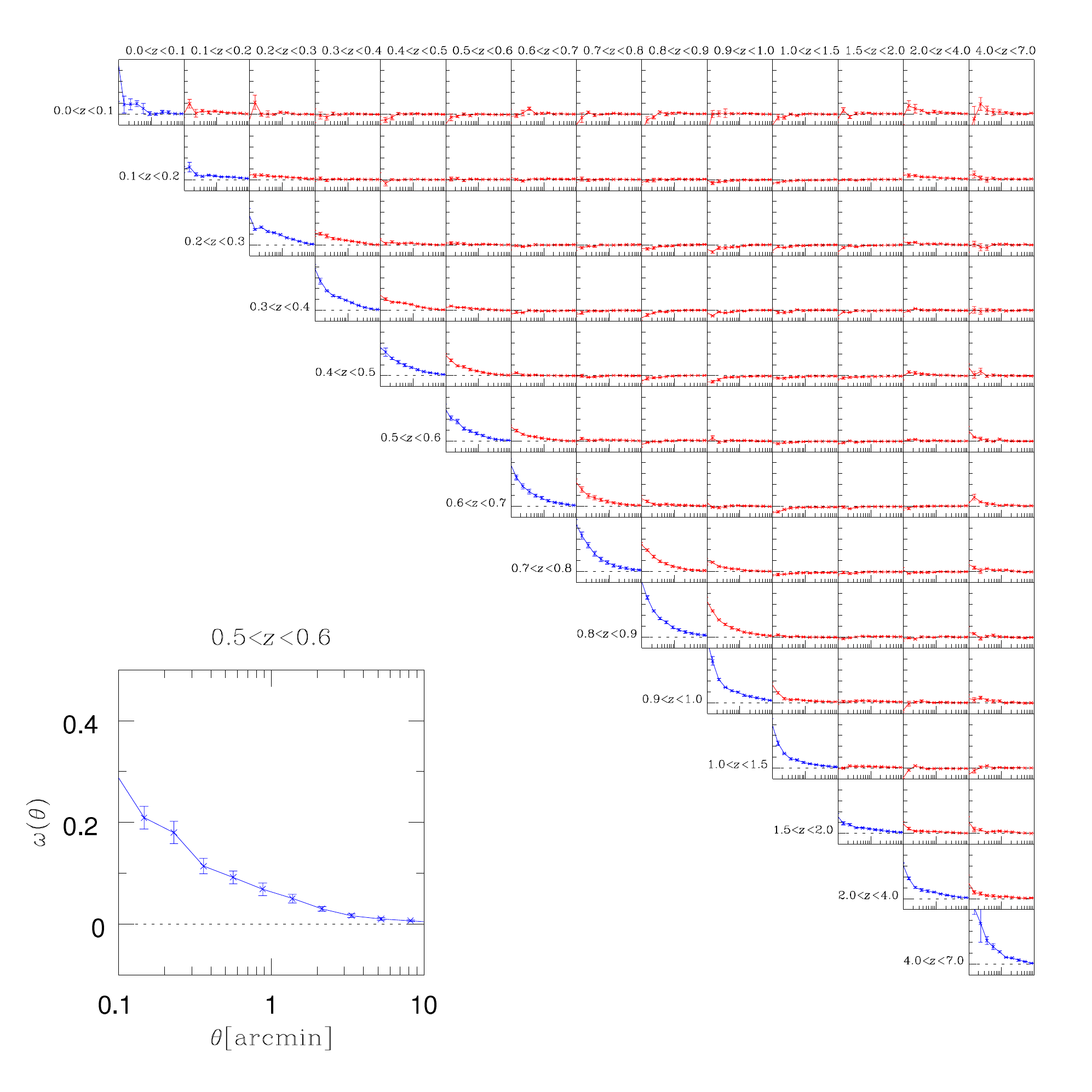}
\caption{ Auto- (\emph{blue} on the diagonal) and cross-correlation
  (\emph{red} off-diagonal) functions of galaxies with $i<26.5$ in
  different photo-$z$ slices. The blow-up in the \emph{bottom left} is
  just an example that illustrates the ranges of the axis.}
\label{fig:cross}
\end{figure*}

Similar to \cite{2009A&A...493.1197E} we estimate the
cross-correlation function of galaxies in photo-$z$ slices. In this
way the null hypothesis of non-overlapping slices, which are far apart
in redshift, can be tested. See \cite{2009A&A...493.1197E} for a
detailed description of the technique. In Fig.~\ref{fig:cross} the
auto- and cross-correlation functions of galaxies with $i<26.5$ in
many different photo-$z$ slices are shown. The fact that nearly all
cross-correlation functions of non-neighbouring photo-$z$ slices show
an amplitude consistent or very close to zero is a strong argument for
the robustness of the photo-$z$'s. No excessive overlap between low-
and high-$z$ slices is observed. Thus, even for fainter magnitudes,
the outlier rates are under control. A more quantitative analysis of
this method will be presented in a forthcoming paper (Benjamin et
al. 2009 in prep.).

\end{document}